\documentclass[%
 aip,
 amsmath,amssymb,
 reprint,%
]{revtex4-1}

\usepackage{graphicx}
\usepackage{dcolumn}
\usepackage{bm}

\usepackage[utf8]{inputenc}
\usepackage[T1]{fontenc}
\usepackage{mathptmx}
\usepackage{float}
\usepackage[final]{changes}

\draft 

\begin{document}


\title{Unraveling radial dependency effects in fiber thermal drawing} 



\author{Alexis G. Page}
\altaffiliation{Alexis G. Page and Mathias Bechert contributed equally to this work.}
\affiliation{Laboratory of Photonic Materials and Fibre Devices, \'Ecole Polytechnique F\'ed\'erale de Lausanne, 1015 Lausanne, Switzerland}
\author{Mathias Bechert}
\altaffiliation{Alexis G. Page and Mathias Bechert contributed equally to this work.}
\author{Fran\c cois Gallaire}
\affiliation{Laboratory of Fluid Mechanics and Instabilities, \'Ecole Polytechnique F\'ed\'erale de Lausanne, 1015 Lausanne, Switzerland}
\author{Fabien Sorin}
 \email{fabien.sorin@epfl.ch}

\affiliation{Laboratory of Photonic Materials and Fibre Devices, \'Ecole Polytechnique F\'ed\'erale de Lausanne, 1015 Lausanne, Switzerland}





\date{\today}


\begin{abstract}

Fiber-based devices with advanced functionalities are emerging as promising solutions for various applications in flexible electronics and bioengineering.  Multi-material thermal drawing in particular has attracted a strong interest for its ability to generate fibers with complex architectures. Thus far however, the understanding of its fluid dynamics has only been applied to single material preforms for which higher order effects, such as the radial dependency of the axial velocity, could be neglected. With \replaced{complex}{novel} multi-material preforms, such effects must be taken into account, as they can affect the architecture and the functional properties of the resulting fiber device. Here, we propose a versatile model of the thermal drawing of fibers which takes into account a radially varying axial velocity.  Unlike the commonly used cross-section averaged approach, our model is capable of predicting radial variations of functional properties caused by the deformation during drawing. This is demonstrated for two effects observed, namely by unraveling the deformation of initially straight, transversal lines in the preform and the dependence on the draw ratio and radial position of the in-fiber electrical conductivity of polymer nano-composites, an important class of materials for emerging fiber devices. This work sets a thus far missing theoretical and practical understanding of multi-material fiber processing to better engineer advanced fibers and textiles for sensing, health care, robotics or bioengineering applications.
\end{abstract}

\pacs{}

\maketitle 



%
%

%

\begin{figure}
\includegraphics[width=\columnwidth]{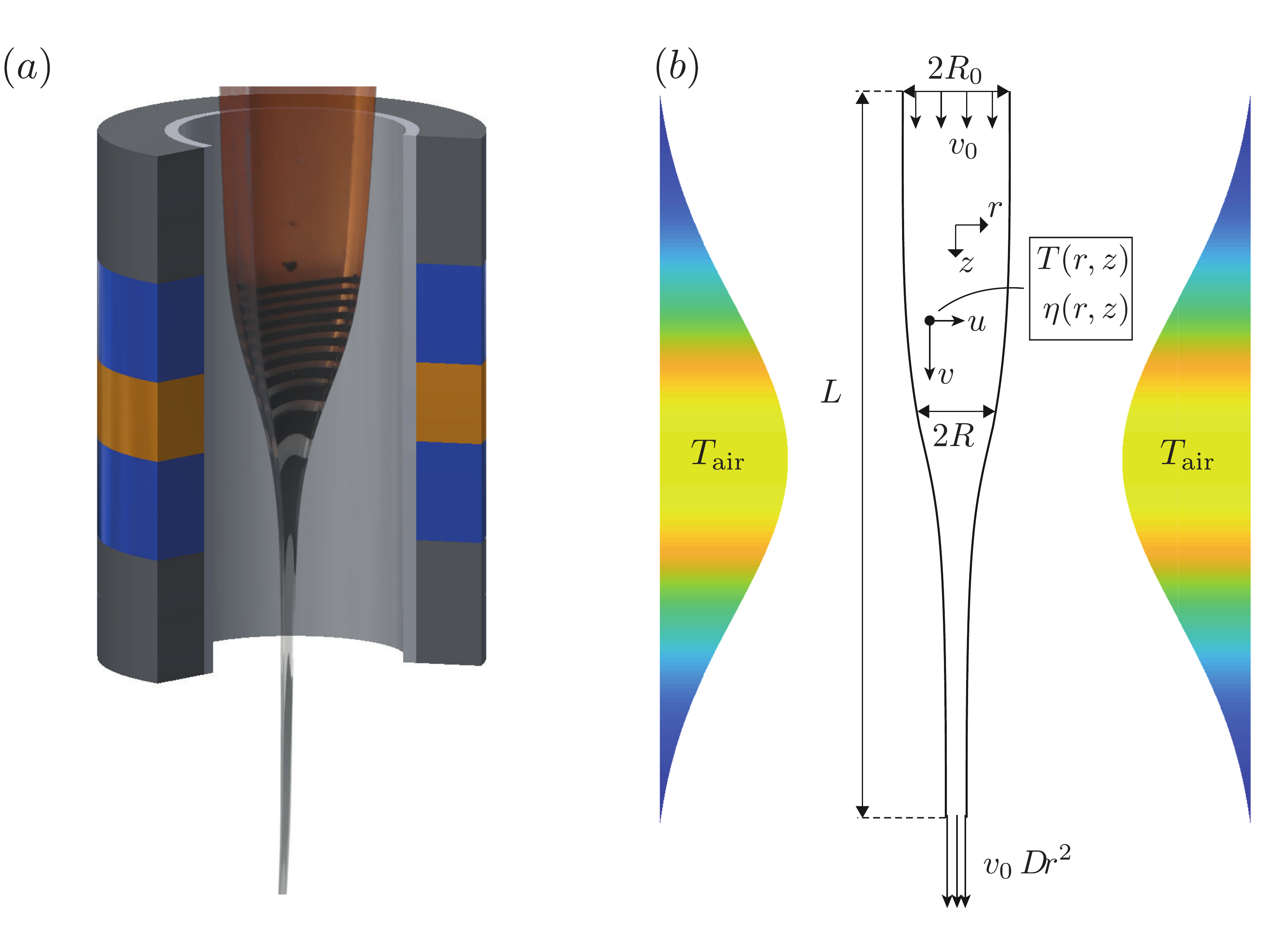}
  \caption{(a) Visualization of the thermal drawing of a fiber in a furnace with three heating zones. We combined a picture of an interrupted draw with a schematic of the furnace. (b)~Sketch of the axisymmetric model. The ambient temperature $T_{\rm air}$ is assumed to have a Gaussian shape along the $z$-axis.}
  \label{fig:drawing_process}
\end{figure}

Thermal drawing is at the heart of the fabrication of telecommunication optical fibers. As illustrated in Fig.~\ref{fig:drawing_process}(a,b), it consists in  heating a macroscale preform, typically made out of glass or thermoplastics, to its softening temperature and pulling it to create a much thinner and longer fiber that maintains the initial cross-sectional architecture. From simple glass-based step index structures, the design of optical fibers was expanded to \added{single-mode waveguiding,} photonic crystal and Bragg mirror fibers \added{\cite{kuzyk_guesthost_1991,eijkelenborg_microstructured_2001,yeung_experimental_2004,peng_development_2004,argyros_09,alexander_schmidt_hybrid_2016,yan_2018}}. \replaced{T}{Recently, t}his technique has also been used to make micro-structured multi-material fibers that integrate optical, electronic and optoelectronic materials \added{\cite{welker_fabrication_1998,abouraddy_towards_2007,sorin_resolving_2010,yan_2018,rein_diode_2018}}. These fibers exhibit advanced functionalities \cite{abouraddy_towards_2007,sorin_multimaterial_2007}, making them promising building blocks for applications in optics but also soft electronics \cite{egusa_multimaterial_2010,gu_soft_2010,khudiyev_2017,qu_superelastic_2018}, optoelectronics \cite{egusa_multimaterial_2010,sorin_resolving_2010,stolyarov_2012,yan_2017,rein_diode_2018}, sensing \cite{nguyen-dang_controlled_2017,qu_superelastic_2018} or bioengineering \cite{nguyen-dang_controlled_2017,guo_polymer_2017,chen_2017}. \par

While the introduction of different materials in the drawing process has triggered much interest to realize innovative fiber-based devices and smart textiles, subtle effects of the fluid dynamics of multi-material co-drawing remain to be investigated to better understand and exploit this approach. In particular, modeling and analytical analysis have mostly relied on fluid dynamics descriptions that assume a velocity in the drawing direction that is independent of the radial position \cite{nguyen-dang_controlled_2017,pone_2006,xue_fabrication_2005}. There exist some numerical studies that are not based on such an assumption, but only in the context of single material, micro-structured optical fibers \cite{xue_fabrication_2005,xue_heat_2017}. The impact of the non-uniform velocity field in the cross-section in the context of multi-material drawing is however essential, albeit yet unexplored. A radially varying velocity field and hence deformation rate during drawing can influence the targeted architecture and resulting properties of the functional materials at the fiber level. A striking manifestation of this effect is schematically shown in Fig.~\ref{fig:drawing_process}(a), where initially rectangular domains of polymer nano-composites deform significantly during drawing. The pattern observed, which we experimentally show and model below, results from a relative displacement of material at different radial positions. Taking this effect into account is primordial to realize advanced fiber-based devices with controlled properties.\par

In this work, we report an in-depth fluid dynamics analysis of the thermal drawing process that takes into account the radial dependency of the axial velocity. We propose a model that includes higher order terms of the velocity field, which enables us to elucidate the observation of line deformation  shown in Fig.~\ref{fig:drawing_process}(a). From this velocity field, we can extract a more accurate distribution of the rate of deformation during drawing, which is crucial for many material properties at the fiber level. In particular, we apply the results of our model to precisely account for the dependence of electrical conductivity of polymer nano-composites on the radial position and the draw ratio, i.e. the ratio of preform to fiber diameters. Conductive polymer composites form an important class of materials widely used in fiber-based devices for optoelectronics \cite{stolyarov_2012, sorin_resolving_2010,yan_2017}, bioengineering \cite{guo_polymer_2017,chen_2017}, and electromechanical sensing \cite{khudiyev_2017,nguyen-dang_multi-material_2017}. They however have a well-known, yet so far unexplained dependence of their electrical conductivity on the fiber drawing conditions \cite{gu_soft_2010}. Our model advances the understanding of multi-material fiber drawing and opens \deleted{novel} opportunities for the design of advanced fiber-based devices and smart textiles with optimum control over material microstructure.

We extend the reduced one-dimensional model for elongating fibers widely used in literature \citep{matovich_1969,shah_1972,german_2008,bechert_combined_2017} to cover thermal effects and to access information about the radial distribution of the variables. We consider only the cladding material in the model, as the functional materials represent only a small fraction of the fiber cross-sectional area, and assume an axisymmetric, round fiber. A detailed derivation of the equations is given in the supplementary material. As shown in Fig.~\ref{fig:drawing_process}(b), the fiber can be described by its radius $R(z)$ along the main axis together with the flow velocity field $(u,v)$, with $u(r,z)$ and $v(r,z)$ denoting the radial and axial velocity components, respectively, as well as temperature $T(r,z)$ and shear viscosity $\eta(r,z)$. All variables are made dimensionless using the initial feeding speed $v_0$, the drawing length $L$, maximum air temperature $T_{\rm max}$ and corresponding minimum viscosity $\eta_{\rm min}$ as reference values. Assuming a slender fiber \cite{schultz_1982,bechert_combined_2017}, i.e. $\alpha^2=(R_0/L)^2\ll 1$, enables us to expand the variables in $\alpha^2$, e.g.,
\begin{align}
\label{eq:expansion}
v(r,z) = v^{(0)} + \alpha^2v^{(1)} + \mathcal{O}\left(\alpha^4\right),
\end{align}
with the superscript indicating the order of expansion. Using an effective drawing length of $L=40\,\rm cm$ and an initial radius of $R_0=1.25\,\rm cm$, $\alpha^2 = \mathcal{O}(10^{-3})$.\par

Following a previous study \cite{scheid_stabilization_2009}, we assume a parabolic temperature field in $r$,
\begin{align}
\label{eq:T_quadr}
T^{(0)}(r,z) = \langle T^{(0)}\rangle + Bi\, \frac{\left(\langle T^{(0)}\rangle-T_{\rm air}\right) R}{4+Bi\,R}\left(1-2\,\frac{r^2}{R^2}\right),
\end{align}
which enables us to calculate the cross-section averaged governing equations. The cross-section average is denoted by $\langle\cdot\rangle$, and the Biot number, which compares the heat exchange between the material and the surrounding air to the heat conduction within the fiber, is defined by $Bi = R_0\, h/k$, $h$ being the heat transfer coefficient and $k$ the material heat conductivity. \added{Note that Eq.~(\ref{eq:T_quadr}) implicitly assumes that the preform is primarily heated by the surrounding air, even though weak radiative heating can be accounted for by using this approach as well \cite{scheid_2012}. In some thermal drawing processes radiative heating can lead to a material temperature higher than the air temperature \cite{garvey_single-mode_1996}, which cannot be modeled by this approach. Similarly, some functional materials may change the mechanical and heat properties of the fiber significantly, despite their low 
proportion inside the fiber.}\par

For the description of the surrounding air temperature, we assume a centered Gaussian profile,
\begin{align}
\label{eq:T_prof}
T_{\rm air}(z) = \frac{\Lambda\,\exp\left[-\frac{\left(z-0.5\right)^2}{\Delta^2}\right]+1}{\Lambda+1},
\end{align}
with dimensionless amplitude $\Lambda$ and width $\Delta$.\par

Utilizing the expansion in $\alpha^2$, the steady state leading order equations are obtained by cross-section averaging the continuity, momentum and heat equations, which yields
\begin{subequations}
\label{eq:cont_mom_heat}
\begin{align}
\partial_z \left(R^2\,v^{(0)}\right) &= 0,\\
\label{eq:mom}
\partial_z \left(R^2\,\langle\eta^{(0)}\rangle\,\partial_z v^{(0)}\right) &= 0,\\
\label{eq:heat}
v^{(0)}\partial_z \langle T^{(0)}\rangle &= -St\,\frac{\langle T^{(0)}\rangle-T_{\rm air}}{R(1+Bi\,R/4)},
\end{align}
\end{subequations}
with radially independent axial velocity $v^{(0)} = v^{(0)}(z)$. The Stanton number $St$ compares the heat transfer between the fiber and the surrounding air to the heat advected by the polymer flow and is given by $St = 2\,Bi/(\alpha\,Pe)$, with the Peclet number $Pe=\rho\,c_p\,v_0\,R_0 / k$, $c_p$ denoting the heat capacity of the material. The energy equation~(\ref{eq:heat}) is coupled to the momentum equation~(\ref{eq:mom}) using Arrhenius' law,
\begin{align}
\label{eq:arrhenius}
\eta = \exp\left[\mu\left(\frac{1}{T}-1\right)\right],
\end{align}
with dimensionless activation energy $\mu$. The streamwise boundary conditions are given by the process setup as
\begin{subequations}
\label{eq:BC}
\begin{align}
v^{(0)}(0) = R(0) = 0,\ \ \  \langle T^{(0)}\rangle(0) &= T_{\rm air}(0),\\
v^{(0)} (1) = Dr^2,
\end{align}
\end{subequations}
with the draw ratio $Dr$ defined by the ratio of preform to fiber diameters. $h, \Lambda$ and $\Delta$ are determined using the experimentally determined preform-to-fiber radius profile during drawing. A detailed description of the parameter determination procedure is given in the supplementary material.\par
We solve Eqs~(\ref{eq:cont_mom_heat}-\ref{eq:BC}), which will also be referred to as the \textit{cross-section averaged model} in the following, by numerical continuation using \textsc{auto97} \cite{auto} to obtain $R$, $v^{(0)}$, $u^{(0)}=-r\,\partial_zv^{(0)}/2$, $\langle T^{(0)}\rangle$ and $\langle\eta^{(0)}\rangle$. Using these solutions, we then go beyond the usual analysis by calculating the radially varying $T^{(0)}$ and $\eta^{(0)}$, as well as investigating the $\alpha^2$-order of the expanded governing equations, which yield
\begin{align}
\label{eq:v1}
\partial_r\left[\eta^{(0)}r\left(\partial_rv^{(1)} - \frac{r}{2}\partial_{zz}v^{(0)}\right)\right] = - 3\,r\,\partial_z\left(\eta^{(0)}\partial_zv^{(0)}\right).
\end{align}
Equation~(\ref{eq:v1}) can be solved by numerical integration to obtain $v^{(1)}$ and $u^{(1)}=\int_0^r dr\,(r\,\partial_zv^{(1)})/r$, both depending on $r$ and $z$. \par

\begin{figure}
\centering
  \includegraphics[width=0.9\columnwidth]{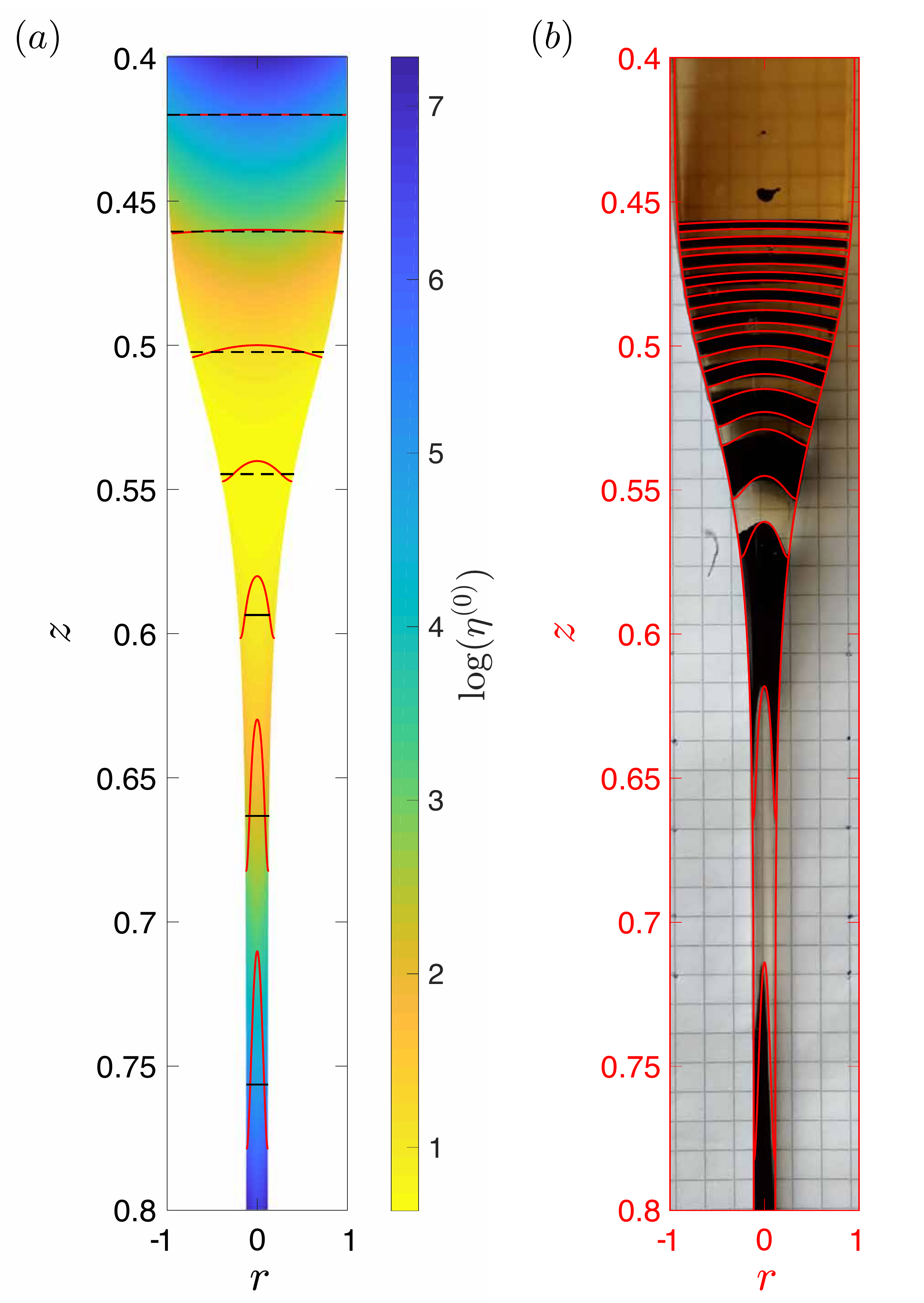}
  \caption{Effect of higher order terms on the evolution of initially straight material lines during the drawing process. (a)~The black dashed lines are calculated with the cross-section averaged model, while the red lines result from the full model. The color map shows the logarithm of the dimensionless viscosity. (b)~Comparison of theoretical prediction (red lines) with experimental observation for a PSu fiber. Parameters: $Dr=8.31,\, Bi = 1,\, St = 32,\, \mu=53,\, \Lambda=0.5,\, \Delta=0.22$.}
 \label{fig:material_lines}
\end{figure}

To demonstrate the importance of this last step, we calculate the evolution of an initially straight line perpendicular to the drawing direction during the drawing process. Fig.~\ref{fig:material_lines}(a) shows the line shape at several axial positions along the fiber within the domain of deformation. Using only the cross-section averaged model (black dashed lines), an initially straight line remains straight during the entire process, which is not surprising as the dominating axial velocity does not vary along the radius at leading order. However, if the higher order terms $u^{(1)}$ and $v^{(1)}$ are included in the calculation (red lines), the lines clearly deform during the drawing, finally looking like an inverted `V'.\par

In order to visualize this effect experimentally, we fabricate (see supplementary material) a polysulfone (PSu) preform of rectangular cross-section ($2.5\,\rm cm\,\times\,1.4\,\rm cm$) within which $1\,\rm{mm}$-wide sheets of carbon-black-loaded polycarbonate (PC) are equidistantly placed, perpendicular to the drawing direction. The feeding speed is set to $2\,\rm{mm/s}$ and the draw ratio is $Dr=8.31$. The furnace has three heating zones (see Fig.~\ref{fig:drawing_process}(a)), which are set to $200/300/120\,^\circ \rm C$. The drawing of this preform is interrupted so that the sheets visualize the deformation of initially straight, horizontal material lines during drawing, as shown in Fig.~\ref{fig:material_lines}(b). \par

The calculated (red) lines coincide very well with the observed shape of the sheets in the PSu fiber. Note that this comparison is nevertheless of qualitative nature, as the fiber used in the experiment has a rectangular cross-section, while the model assumes a circular fiber\replaced{. A}{, and a} PC fiber is found to yield less quantitative, but still qualitative coincidence with the model prediction\added{, which is probably due to the simplification of heating mechanisms as discussed above.} \added{For both materials, however, the final shape of the deformed lines is reproduced with high accuracy.} All parameter values are given in the caption of Fig.~\ref{fig:material_lines}. Physically, the non-uniform propagation of fluid elements lying on an initially straight line is the result of an interplay between a radially varying viscosity (see Fig.~\ref{fig:material_lines}(a) and supplementary material) and the deformation of the fiber due to drawing.\par

It is worth mentioning that the significant influence of the higher order velocity terms is a cumulative effect. At one particular position in $z$, the change in velocity introduced by the higher order is at most only a few percent. This is expected, as the contribution is scaled by $\alpha^2$ according to Eq.~(\ref{eq:expansion}), with $\alpha^2$ being small at the basis of the expansion. By following a material point during the drawing, however, we integrate along the fiber length, and by that the order of magnitude increases from $\alpha^2=(R_0/L)^2$ to $R_0^2/L$. For this reason, the widely used cross-section averaged model, Eqs.(\ref{eq:cont_mom_heat}-\ref{eq:BC}), is well-suited if one is interested exclusively in the fiber radius and the averaged velocity field, as it is the case for single-material fibers, but insufficient if radially distributed properties come into play, like for multi-material fibers.\par
\begin{figure}[htp]
  \includegraphics[width=\columnwidth]{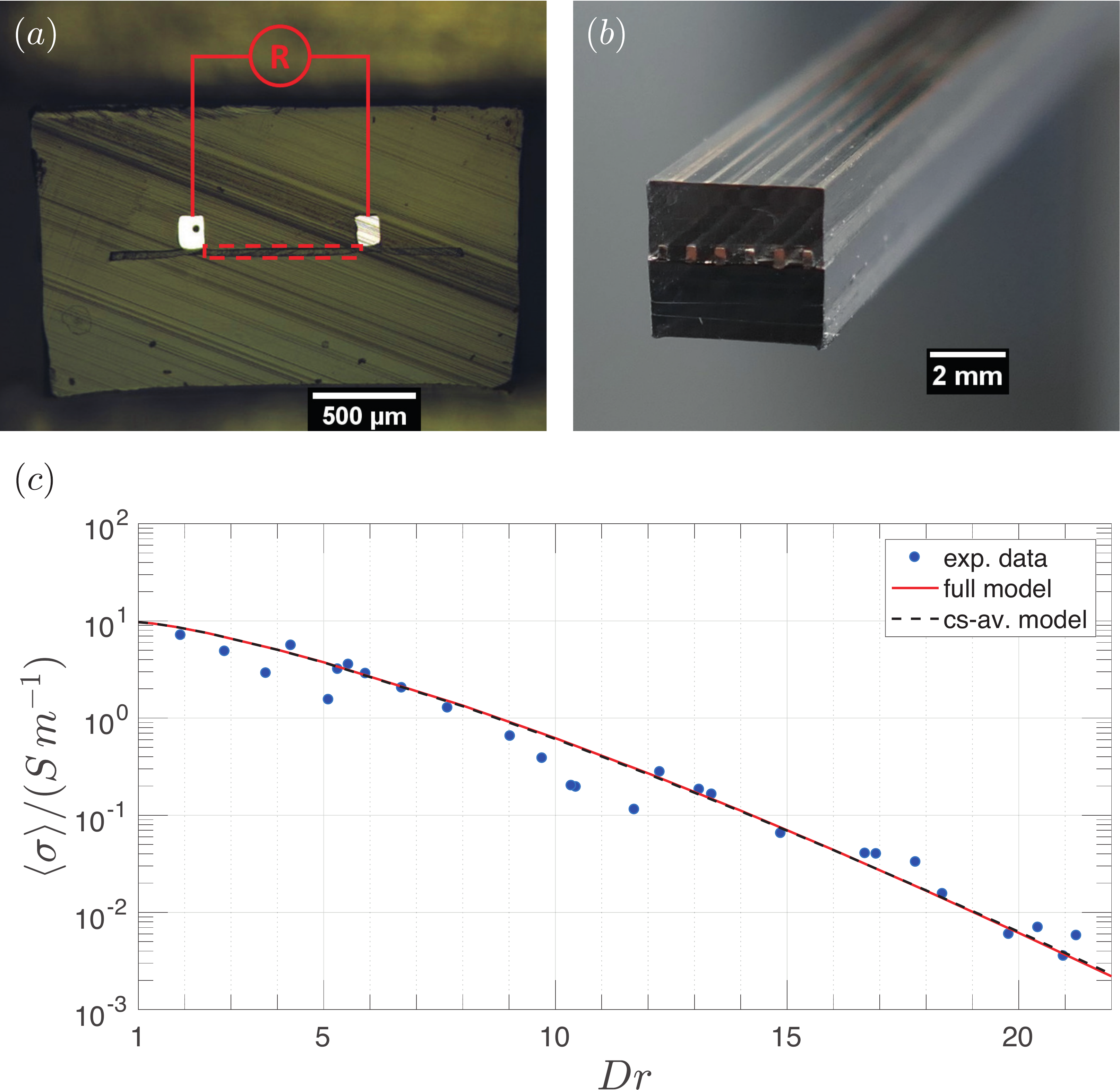}
  \caption{(a)~Optical micrograph of the cross-section of a fiber with PSu cladding, PC-CNT composite layer and two metallic electrodes, which enable to measure the transversal electrical conductivity for the domain marked by the dashed line. (b)~Perspective view of a fiber with PC cladding, PC-CB composite layer and six metallic electrodes distributed in the cross section. (c)~Averaged conductivity of the PC-CNT composite sheet in a multimaterial fiber with a PSu cladding as a function of draw ratio. The perpendicular conductivity is measured between two electrodes placed at $r/R = \pm 0.6$, as shown in (a). Blue points show experimental data while the red and black, dashed lines depict the theoretical prediction of the full and cross-section averaged models, respectively. Parameters: $Bi=1.,\, St = 32,\, \mu = 53,\, \Lambda=0.5,\,\Delta = 0.22,\, m=1.5,\, \sigma_{\rm in} = 9.7\, \rm S\,m^{-1},$ $a\,c = 3.9\times10^{-3}$ (full model), $a\,c = 3.6\times10^{-3}$ (cross-section averaged model).}
  \label{fig:cond_PSu_PCCNT}
\end{figure}

To further highlight the importance of this \deleted{new} understanding, we now investigate the influence of thermal drawing on the electrical conductivity of nano-composites. As conductive fillers, we use carbon black (CB) and carbon nanotubes (CNT). In particular, we fabricate two types of preform: a PSu cladding with a PC-CNT composite sheet and a PC cladding with a PC-CB composite sheet.  For drawing the preform with PSu cladding, we use the same settings as above. The preform with PC cladding has a rectangular cross-section ($2.5\,\rm cm\,\times\,2.3\,\rm cm$), the feeding speed is set to $2\,\rm{mm/s}$, and the heating zones of the furnace are set to $120/270/120\,^\circ \rm C$.\par

As visible in Figs.~\ref{fig:cond_PSu_PCCNT}(a,b), the nano-composite sheet is in contact with several metallic ribbons ($2\,\rm{mm}\times1\,\rm{mm}$ cross-section) made of a bismuth-tin (tin-zinc) alloy for the PC (PSu) cladding. Even though the ribbons melt during consolidation and drawing, the high viscosities of the cladding and sheet materials ensure that the cross-sectional architecture is preserved. By polishing away the cladding of the drawn fiber, we can individually contact the electrodes with a multimeter to measure the conductivity perpendicular to the drawing direction (see Fig.~\ref{fig:cond_PSu_PCCNT}(a)).\par
In order to access theoretically the final conductivity, we use a simple percolation law to link the effective filler concentration $p_e$ to the conductivity $\sigma$,
\begin{align}
\label{eq:percolation}
\sigma = \sigma_{\rm in}\,\left(\frac{p_e-p_c}{p_{\rm in}-p_c}\right)^c,
\end{align}
with initial conductivity $\sigma_{\rm in}$ of the undeformed material having an initial filler concentration $p_{\rm in}$, critical concentration $p_c$, and critical exponent $c$. To account for the destruction of conductive pathways due to deformation during drawing we propose an empirical kinetic equation inspired by a previous work \cite{saphiannikova_superposition_2012}, which links the effective filler concentration to the second invariant of the deformation tensor $\sqrt{II_D}$ (see supplementary material),
\begin{align}
\label{eq:kinetic}
\partial_t p_e = -a\,\eta\,\sqrt{II_D}^{\,m}(p_e-p_c).
\end{align}
$a$ denotes a destruction rate and analogue to the previous work \cite{saphiannikova_superposition_2012}, we set $m=1.5$. Solving Eq.~(\ref{eq:kinetic}) and substituting in (\ref{eq:percolation}) then yields
\begin{align}
\label{eq:cond_distr}
\sigma(r_f) &= \sigma_{\rm in}\exp\left(-a\,c\int_{\gamma(r_f)}\frac{dz}{v}\,\sqrt{II_D}^{\,m}\eta\right),
\end{align}
with $\gamma(r_f)$ denoting the trajectory of a material element during drawing with final radial position $r_f$. The product $a\,c$ and $\sigma_{\rm in}$ are determined by fitting the experimental data. Equation~(\ref{eq:cond_distr}) yields the radial distribution of conductivity in the drawn fiber, which can then be used to calculate the averaged conductivities measured between two electrodes (see Fig.~\ref{fig:cond_PSu_PCCNT}(a)). \par
Fig.~\ref{fig:cond_PSu_PCCNT}(c) shows the perpendicular conductivity of the PC-CNT composite in a PSu fiber, measured between two electrodes placed at $r/R = \pm0.6$ for a large range of draw ratios, and compares it to the theoretically determined averaged conductivity (red line). The increasing deformation with increasing draw ratio leads to a change in conductivity over several orders of magnitude, which is well reproduced by our model. Also plotted is the theoretical prediction based on the cross-section averaged model (black dashed line). Both models lead to practically identical results, which is not surprising as we average the conductivity over a large part of the fiber radius.\par
\begin{figure}[tb]
  \includegraphics[width=0.85\columnwidth]{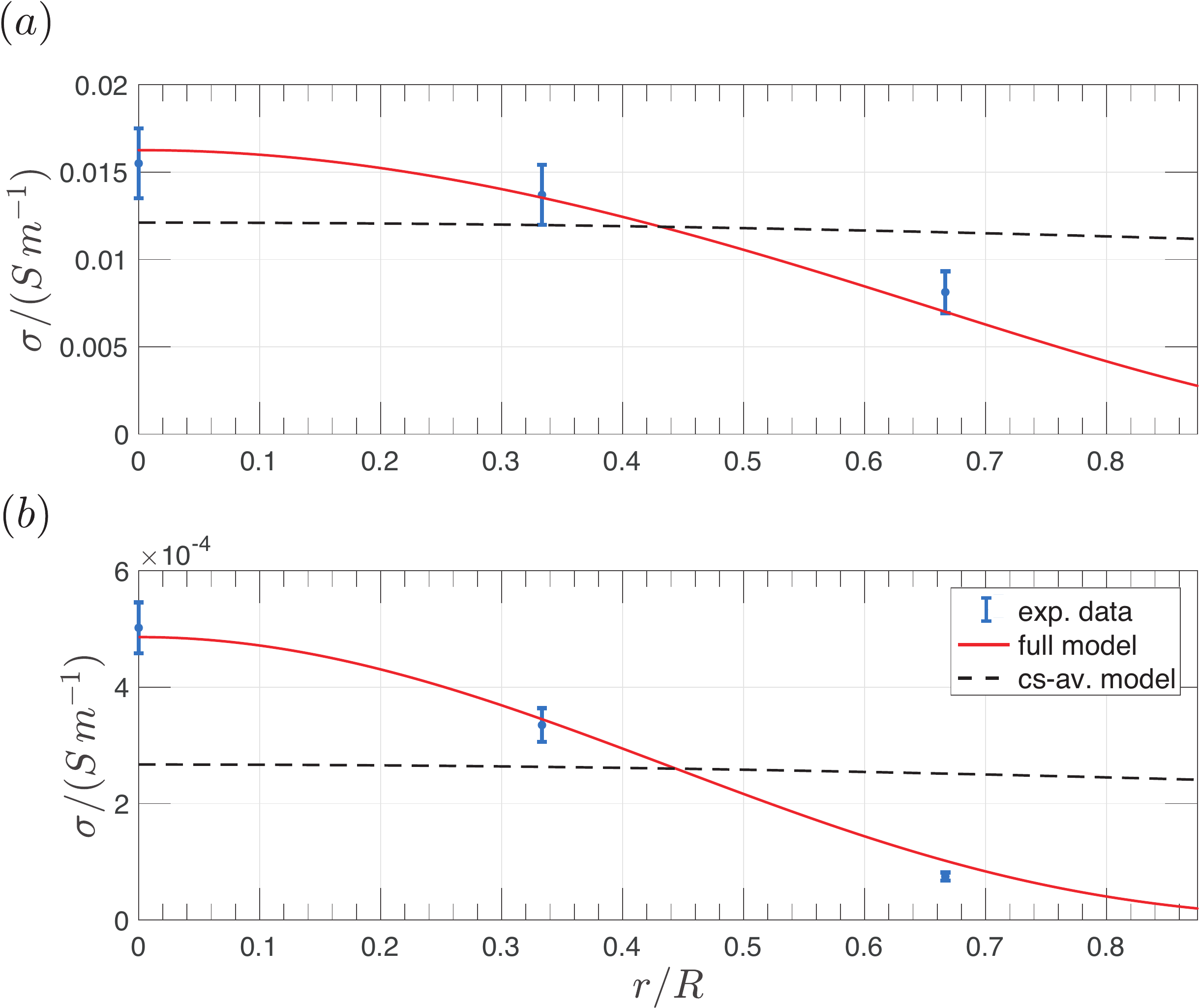}
  \caption{Dependence of the conductivity on the fiber radius for PC-CB composite sheet in a multimaterial fiber with a PC cladding for two draw ratios, (a) $Dr=5.12\pm0.16$ and (b) $Dr=7.24\pm0.13$. The blue points depict the experimentally determined \deleted{averaged} conductivities \added{calculated by averaging 3-10 measurements of the perpendicular conductivity between two metallic electrodes for each point. The error bars correspond to the standard deviation.} The red line shows the theoretically predicted radial distribution of the averaged conductivity and the black, dashed line depicts the distribution obtained using only the cross-section averaged model. Parameters: $Bi=1.5,\, St = 27,\, \mu = 48,\, \Lambda = 0.45,\, \Delta = 0.17,\, m=1.5,\, \sigma_{\rm in} = 2\,\rm S\,m^{-1},\, a\,c = 1.1\times10^{-2}$ (full model), $a\,c = 9.8\times10^{-3}$ (cross-section averaged model).}
  \label{fig:cond_PC_PCCB}
\end{figure}
In order to access experimentally the conductivity as a function of the radial position, we prepare a PC/PC-CB preform with six electrodes placed at equal distances along the fiber width (see Fig.~\ref{fig:cond_PSu_PCCNT}(b)). Fig.~\ref{fig:cond_PC_PCCB} shows the measured conductivities for two different draw ratios. The points are placed in the middle of the domain of average, which has a size of $2.5\,\rm{mm}$, i.e. $0.25\,R$, corresponding to the distance between the electrodes. These measurements are compared to the theoretically obtained moving-averaged distribution (red line). Despite its simple approach, our model describes the variation of conductivity along the radial position very well. Note that analogue to the PSu/PC-CNT fiber, the values of $\sigma_{\rm in}$ and $a\,c$ are uniquely set, independent of the draw ratio. Also shown in Fig.~\ref{fig:cond_PC_PCCB} is the averaged conductivity distribution resulting from the cross-section averaged model (black dashed line). Clearly, this model is incapable of predicting such a large radial variation of the conductivity, emphasizing the need of our extended model. \par

The model presented here is able to predict the radial distribution of material properties in multimaterial fibers fabricated by \added{weakly radiative} thermal drawing, while it maintains the simplicity and flexibility of the cross-section averaged one-dimensional model for single-material fibers. Even though assumptions like using a circular cross-section or employing the simple empirical kinetic equation~(\ref{eq:kinetic}) appear to be rather strong, we are able to reach qualitative and quantitative coincidence with our experimental findings. The flexibility of the model enables comprehensive parametric studies to find routes to \replaced{innovative}{new} experimental designs. Our approach improves the understanding of multi-material fiber drawing and opens \deleted{novel} opportunities for the design of advanced fiber-based devices with optimum control over material microstructure and tailored properties.

\section*{Supplementary Material}
\added{
The supplemental materials provide a detailed derivation of the thermal drawing model, a description of the materials and experimental methods used in this work, all parameter values used in the modeling, together with a description of how they are obtained, and further information on the evolution of material temperature, viscosity and velocity fields during drawing.
}

%


\begin{thebibliography}{35}%
\makeatletter
\providecommand \@ifxundefined [1]{%
 \@ifx{#1\undefined}
}%
\providecommand \@ifnum [1]{%
 \ifnum #1\expandafter \@firstoftwo
 \else \expandafter \@secondoftwo
 \fi
}%
\providecommand \@ifx [1]{%
 \ifx #1\expandafter \@firstoftwo
 \else \expandafter \@secondoftwo
 \fi
}%
\providecommand \natexlab [1]{#1}%
\providecommand \enquote  [1]{``#1''}%
\providecommand \bibnamefont  [1]{#1}%
\providecommand \bibfnamefont [1]{#1}%
\providecommand \citenamefont [1]{#1}%
\providecommand \href@noop [0]{\@secondoftwo}%
\providecommand \href [0]{\begingroup \@sanitize@url \@href}%
\providecommand \@href[1]{\@@startlink{#1}\@@href}%
\providecommand \@@href[1]{\endgroup#1\@@endlink}%
\providecommand \@sanitize@url [0]{\catcode `\\12\catcode `\$12\catcode
  `\&12\catcode `\#12\catcode `\^12\catcode `\_12\catcode `\%12\relax}%
\providecommand \@@startlink[1]{}%
\providecommand \@@endlink[0]{}%
\providecommand \url  [0]{\begingroup\@sanitize@url \@url }%
\providecommand \@url [1]{\endgroup\@href {#1}{\urlprefix }}%
\providecommand \urlprefix  [0]{URL }%
\providecommand \Eprint [0]{\href }%
\providecommand \doibase [0]{http://dx.doi.org/}%
\providecommand \selectlanguage [0]{\@gobble}%
\providecommand \bibinfo  [0]{\@secondoftwo}%
\providecommand \bibfield  [0]{\@secondoftwo}%
\providecommand \translation [1]{[#1]}%
\providecommand \BibitemOpen [0]{}%
\providecommand \bibitemStop [0]{}%
\providecommand \bibitemNoStop [0]{.\EOS\space}%
\providecommand \EOS [0]{\spacefactor3000\relax}%
\providecommand \BibitemShut  [1]{\csname bibitem#1\endcsname}%
\let\auto@bib@innerbib\@empty
\bibitem [{\citenamefont {Kuzyk}, \citenamefont {Paek},\ and\ \citenamefont
  {Dirk}(1991)}]{kuzyk_guesthost_1991}%
  \BibitemOpen
  \bibfield  {author} {\bibinfo {author} {\bibfnamefont {M.~G.}\ \bibnamefont
  {Kuzyk}}, \bibinfo {author} {\bibfnamefont {U.~C.}\ \bibnamefont {Paek}}, \
  and\ \bibinfo {author} {\bibfnamefont {C.~W.}\ \bibnamefont {Dirk}},\
  }\bibfield  {title} {\enquote {\bibinfo {title} {Guest‐host polymer fibers
  for nonlinear optics},}\ }\href {\doibase 10.1063/1.105271} {\bibfield
  {journal} {\bibinfo  {journal} {Applied Physics Letters}\ }\textbf {\bibinfo
  {volume} {59}},\ \bibinfo {pages} {902--904} (\bibinfo {year}
  {1991})}\BibitemShut {NoStop}%
\bibitem [{\citenamefont {Eijkelenborg}\ \emph {et~al.}(2001)\citenamefont
  {Eijkelenborg}, \citenamefont {Large}, \citenamefont {Argyros}, \citenamefont
  {Zagari}, \citenamefont {Manos}, \citenamefont {Issa}, \citenamefont
  {Bassett}, \citenamefont {Fleming}, \citenamefont {McPhedran}, \citenamefont
  {Sterke},\ and\ \citenamefont
  {Nicorovici}}]{eijkelenborg_microstructured_2001}%
  \BibitemOpen
  \bibfield  {author} {\bibinfo {author} {\bibfnamefont {M.~A.~v.}\
  \bibnamefont {Eijkelenborg}}, \bibinfo {author} {\bibfnamefont {M.~C.~J.}\
  \bibnamefont {Large}}, \bibinfo {author} {\bibfnamefont {A.}~\bibnamefont
  {Argyros}}, \bibinfo {author} {\bibfnamefont {J.}~\bibnamefont {Zagari}},
  \bibinfo {author} {\bibfnamefont {S.}~\bibnamefont {Manos}}, \bibinfo
  {author} {\bibfnamefont {N.~A.}\ \bibnamefont {Issa}}, \bibinfo {author}
  {\bibfnamefont {I.}~\bibnamefont {Bassett}}, \bibinfo {author} {\bibfnamefont
  {S.}~\bibnamefont {Fleming}}, \bibinfo {author} {\bibfnamefont {R.~C.}\
  \bibnamefont {McPhedran}}, \bibinfo {author} {\bibfnamefont {C.~M.~d.}\
  \bibnamefont {Sterke}}, \ and\ \bibinfo {author} {\bibfnamefont {N.~A.~P.}\
  \bibnamefont {Nicorovici}},\ }\bibfield  {title} {\enquote {\bibinfo {title}
  {Microstructured polymer optical fibre},}\ }\href {\doibase
  10.1364/OE.9.000319} {\bibfield  {journal} {\bibinfo  {journal} {Optics
  Express}\ }\textbf {\bibinfo {volume} {9}},\ \bibinfo {pages} {319--327}
  (\bibinfo {year} {2001})}\BibitemShut {NoStop}%
\bibitem [{\citenamefont {Yeung}\ \emph {et~al.}(2004)\citenamefont {Yeung},
  \citenamefont {Chiang}, \citenamefont {Rastogi}, \citenamefont {Chu},\ and\
  \citenamefont {Peng}}]{yeung_experimental_2004}%
  \BibitemOpen
  \bibfield  {author} {\bibinfo {author} {\bibfnamefont {A.}~\bibnamefont
  {Yeung}}, \bibinfo {author} {\bibfnamefont {K.}~\bibnamefont {Chiang}},
  \bibinfo {author} {\bibfnamefont {V.}~\bibnamefont {Rastogi}}, \bibinfo
  {author} {\bibfnamefont {P.}~\bibnamefont {Chu}}, \ and\ \bibinfo {author}
  {\bibfnamefont {G.}~\bibnamefont {Peng}},\ }\bibfield  {title} {\enquote
  {\bibinfo {title} {Experimental demonstration of single-mode operation of
  large-core segmented cladding fiber},}\ }in\ \href@noop {} {\emph {\bibinfo
  {booktitle} {Optical Fiber Communication Conference. OFC 2004}}},\
  Vol.~\bibinfo {volume} {2}\ (\bibinfo {organization} {IEEE},\ \bibinfo {year}
  {2004})\BibitemShut {NoStop}%
\bibitem [{\citenamefont {Peng}, \citenamefont {Ji},\ and\ \citenamefont
  {Wang}(2004)}]{peng_development_2004}%
  \BibitemOpen
  \bibfield  {author} {\bibinfo {author} {\bibfnamefont {G.-D.}\ \bibnamefont
  {Peng}}, \bibinfo {author} {\bibfnamefont {P.~N.}\ \bibnamefont {Ji}}, \ and\
  \bibinfo {author} {\bibfnamefont {T.}~\bibnamefont {Wang}},\ }\bibfield
  {title} {\enquote {\bibinfo {title} {Development of special polymer optical
  fibers and devices},}\ }in\ \href@noop {} {\emph {\bibinfo {booktitle}
  {Active and {Passive} {Optical} {Components} for {WDM} {Communications}
  {IV}}}},\ Vol.\ \bibinfo {volume} {5595}\ (\bibinfo  {publisher}
  {International Society for Optics and Photonics},\ \bibinfo {year} {2004})\
  pp.\ \bibinfo {pages} {138--153}\BibitemShut {NoStop}%
\bibitem [{\citenamefont {Argyros}(2009)}]{argyros_09}%
  \BibitemOpen
  \bibfield  {author} {\bibinfo {author} {\bibfnamefont {A.}~\bibnamefont
  {Argyros}},\ }\bibfield  {title} {\enquote {\bibinfo {title} {Microstructured
  polymer optical fibers},}\ }\href
  {http://jlt.osa.org/abstract.cfm?URI=jlt-27-11-1571} {\bibfield  {journal}
  {\bibinfo  {journal} {J. Lightwave Technol.}\ }\textbf {\bibinfo {volume}
  {27}},\ \bibinfo {pages} {1571--1579} (\bibinfo {year} {2009})}\BibitemShut
  {NoStop}%
\bibitem [{\citenamefont {Alexander~Schmidt}, \citenamefont {Argyros},\ and\
  \citenamefont {Sorin}(2016)}]{alexander_schmidt_hybrid_2016}%
  \BibitemOpen
  \bibfield  {author} {\bibinfo {author} {\bibfnamefont {M.}~\bibnamefont
  {Alexander~Schmidt}}, \bibinfo {author} {\bibfnamefont {A.}~\bibnamefont
  {Argyros}}, \ and\ \bibinfo {author} {\bibfnamefont {F.}~\bibnamefont
  {Sorin}},\ }\bibfield  {title} {\enquote {\bibinfo {title} {Hybrid {Optical}
  {Fibers} - {An} {Innovative} {Platform} for {In}-{Fiber} {Photonic}
  {Devices}},}\ }\href {\doibase 10.1002/adom.201500319} {\bibfield  {journal}
  {\bibinfo  {journal} {Advanced Optical Materials}\ }\textbf {\bibinfo
  {volume} {4}},\ \bibinfo {pages} {13--36} (\bibinfo {year}
  {2016})}\BibitemShut {NoStop}%
\bibitem [{\citenamefont {Yan}\ \emph {et~al.}(2018)\citenamefont {Yan},
  \citenamefont {Page}, \citenamefont {Nguyen-Dang}, \citenamefont {Qu},
  \citenamefont {Sordo}, \citenamefont {Wei},\ and\ \citenamefont
  {Sorin}}]{yan_2018}%
  \BibitemOpen
  \bibfield  {author} {\bibinfo {author} {\bibfnamefont {W.}~\bibnamefont
  {Yan}}, \bibinfo {author} {\bibfnamefont {A.}~\bibnamefont {Page}}, \bibinfo
  {author} {\bibfnamefont {T.}~\bibnamefont {Nguyen-Dang}}, \bibinfo {author}
  {\bibfnamefont {Y.}~\bibnamefont {Qu}}, \bibinfo {author} {\bibfnamefont
  {F.}~\bibnamefont {Sordo}}, \bibinfo {author} {\bibfnamefont
  {L.}~\bibnamefont {Wei}}, \ and\ \bibinfo {author} {\bibfnamefont
  {F.}~\bibnamefont {Sorin}},\ }\bibfield  {title} {\enquote {\bibinfo {title}
  {Advanced multimaterial electronic and optoelectronic fibers and textiles},}\
  }\href@noop {} {\bibfield  {journal} {\bibinfo  {journal} {Advanced
  Materials}\ }\textbf {\bibinfo {volume} {31}},\ \bibinfo {pages} {1802348}
  (\bibinfo {year} {2018})}\BibitemShut {NoStop}%
\bibitem [{\citenamefont {Welker}\ \emph {et~al.}(1998)\citenamefont {Welker},
  \citenamefont {Tostenrude}, \citenamefont {Garvey}, \citenamefont
  {Canfield},\ and\ \citenamefont {Kuzyk}}]{welker_fabrication_1998}%
  \BibitemOpen
  \bibfield  {author} {\bibinfo {author} {\bibfnamefont {D.~J.}\ \bibnamefont
  {Welker}}, \bibinfo {author} {\bibfnamefont {J.}~\bibnamefont {Tostenrude}},
  \bibinfo {author} {\bibfnamefont {D.~W.}\ \bibnamefont {Garvey}}, \bibinfo
  {author} {\bibfnamefont {B.~K.}\ \bibnamefont {Canfield}}, \ and\ \bibinfo
  {author} {\bibfnamefont {M.~G.}\ \bibnamefont {Kuzyk}},\ }\bibfield  {title}
  {\enquote {\bibinfo {title} {Fabrication and characterization of single-mode
  electro-optic polymer optical fiber},}\ }\href {\doibase
  10.1364/OL.23.001826} {\bibfield  {journal} {\bibinfo  {journal} {Optics
  Letters}\ }\textbf {\bibinfo {volume} {23}},\ \bibinfo {pages} {1826--1828}
  (\bibinfo {year} {1998})}\BibitemShut {NoStop}%
\bibitem [{\citenamefont {Abouraddy}\ \emph {et~al.}(2007)\citenamefont
  {Abouraddy}, \citenamefont {Bayindir}, \citenamefont {Benoit}, \citenamefont
  {Hart}, \citenamefont {Kuriki}, \citenamefont {Orf}, \citenamefont {Shapira},
  \citenamefont {Sorin}, \citenamefont {Temelkuran},\ and\ \citenamefont
  {Fink}}]{abouraddy_towards_2007}%
  \BibitemOpen
  \bibfield  {author} {\bibinfo {author} {\bibfnamefont {A.~F.}\ \bibnamefont
  {Abouraddy}}, \bibinfo {author} {\bibfnamefont {M.}~\bibnamefont {Bayindir}},
  \bibinfo {author} {\bibfnamefont {G.}~\bibnamefont {Benoit}}, \bibinfo
  {author} {\bibfnamefont {S.~D.}\ \bibnamefont {Hart}}, \bibinfo {author}
  {\bibfnamefont {K.}~\bibnamefont {Kuriki}}, \bibinfo {author} {\bibfnamefont
  {N.}~\bibnamefont {Orf}}, \bibinfo {author} {\bibfnamefont {O.}~\bibnamefont
  {Shapira}}, \bibinfo {author} {\bibfnamefont {F.}~\bibnamefont {Sorin}},
  \bibinfo {author} {\bibfnamefont {B.}~\bibnamefont {Temelkuran}}, \ and\
  \bibinfo {author} {\bibfnamefont {Y.}~\bibnamefont {Fink}},\ }\bibfield
  {title} {\enquote {\bibinfo {title} {Towards multimaterial multifunctional
  fibres that see, hear, sense and communicate},}\ }\href {\doibase
  10.1038/nmat1889} {\bibfield  {journal} {\bibinfo  {journal} {Nature
  Materials}\ }\textbf {\bibinfo {volume} {6}},\ \bibinfo {pages} {336--347}
  (\bibinfo {year} {2007})}\BibitemShut {NoStop}%
\bibitem [{\citenamefont {Sorin}\ \emph {et~al.}(2010)\citenamefont {Sorin},
  \citenamefont {Lestoquoy}, \citenamefont {Danto}, \citenamefont
  {Joannopoulos},\ and\ \citenamefont {Fink}}]{sorin_resolving_2010}%
  \BibitemOpen
  \bibfield  {author} {\bibinfo {author} {\bibfnamefont {F.}~\bibnamefont
  {Sorin}}, \bibinfo {author} {\bibfnamefont {G.}~\bibnamefont {Lestoquoy}},
  \bibinfo {author} {\bibfnamefont {S.}~\bibnamefont {Danto}}, \bibinfo
  {author} {\bibfnamefont {J.~D.}\ \bibnamefont {Joannopoulos}}, \ and\
  \bibinfo {author} {\bibfnamefont {Y.}~\bibnamefont {Fink}},\ }\bibfield
  {title} {\enquote {\bibinfo {title} {Resolving optical illumination
  distributions along an axially symmetric photodetecting fiber},}\ }\href
  {\doibase 10.1364/OE.18.024264} {\bibfield  {journal} {\bibinfo  {journal}
  {Optics Express}\ }\textbf {\bibinfo {volume} {18}},\ \bibinfo {pages}
  {24264--24275} (\bibinfo {year} {2010})}\BibitemShut {NoStop}%
\bibitem [{\citenamefont {Rein}\ \emph {et~al.}(2018)\citenamefont {Rein},
  \citenamefont {Favrod}, \citenamefont {Hou}, \citenamefont {Khudiyev},
  \citenamefont {Stolyarov}, \citenamefont {Cox}, \citenamefont {Chung},
  \citenamefont {Chhav}, \citenamefont {Ellis}, \citenamefont {Joannopoulos}
  \emph {et~al.}}]{rein_diode_2018}%
  \BibitemOpen
  \bibfield  {author} {\bibinfo {author} {\bibfnamefont {M.}~\bibnamefont
  {Rein}}, \bibinfo {author} {\bibfnamefont {V.~D.}\ \bibnamefont {Favrod}},
  \bibinfo {author} {\bibfnamefont {C.}~\bibnamefont {Hou}}, \bibinfo {author}
  {\bibfnamefont {T.}~\bibnamefont {Khudiyev}}, \bibinfo {author}
  {\bibfnamefont {A.}~\bibnamefont {Stolyarov}}, \bibinfo {author}
  {\bibfnamefont {J.}~\bibnamefont {Cox}}, \bibinfo {author} {\bibfnamefont
  {C.-C.}\ \bibnamefont {Chung}}, \bibinfo {author} {\bibfnamefont
  {C.}~\bibnamefont {Chhav}}, \bibinfo {author} {\bibfnamefont
  {M.}~\bibnamefont {Ellis}}, \bibinfo {author} {\bibfnamefont
  {J.}~\bibnamefont {Joannopoulos}},  \emph {et~al.},\ }\bibfield  {title}
  {\enquote {\bibinfo {title} {Diode fibres for fabric-based optical
  communications},}\ }\href@noop {} {\bibfield  {journal} {\bibinfo  {journal}
  {Nature}\ }\textbf {\bibinfo {volume} {560}},\ \bibinfo {pages} {214}
  (\bibinfo {year} {2018})}\BibitemShut {NoStop}%
\bibitem [{\citenamefont {Sorin}\ \emph {et~al.}(2007)\citenamefont {Sorin},
  \citenamefont {Abouraddy}, \citenamefont {Orf}, \citenamefont {Shapira},
  \citenamefont {Viens}, \citenamefont {Arnold}, \citenamefont {Joannopoulos},\
  and\ \citenamefont {Fink}}]{sorin_multimaterial_2007}%
  \BibitemOpen
  \bibfield  {author} {\bibinfo {author} {\bibfnamefont {F.}~\bibnamefont
  {Sorin}}, \bibinfo {author} {\bibfnamefont {A.~F.}\ \bibnamefont
  {Abouraddy}}, \bibinfo {author} {\bibfnamefont {N.}~\bibnamefont {Orf}},
  \bibinfo {author} {\bibfnamefont {O.}~\bibnamefont {Shapira}}, \bibinfo
  {author} {\bibfnamefont {J.}~\bibnamefont {Viens}}, \bibinfo {author}
  {\bibfnamefont {J.}~\bibnamefont {Arnold}}, \bibinfo {author} {\bibfnamefont
  {J.~D.}\ \bibnamefont {Joannopoulos}}, \ and\ \bibinfo {author}
  {\bibfnamefont {Y.}~\bibnamefont {Fink}},\ }\bibfield  {title} {\enquote
  {\bibinfo {title} {Multimaterial {Photodetecting} {Fibers}: a {Geometric} and
  {Structural} {Study}},}\ }\href {\doibase 10.1002/adma.200700177} {\bibfield
  {journal} {\bibinfo  {journal} {Advanced Materials}\ }\textbf {\bibinfo
  {volume} {19}},\ \bibinfo {pages} {3872--3877} (\bibinfo {year}
  {2007})}\BibitemShut {NoStop}%
\bibitem [{\citenamefont {Egusa}\ \emph {et~al.}(2010)\citenamefont {Egusa},
  \citenamefont {Wang}, \citenamefont {Chocat}, \citenamefont {Ruff},
  \citenamefont {Stolyarov}, \citenamefont {Shemuly}, \citenamefont {Sorin},
  \citenamefont {Rakich}, \citenamefont {Joannopoulos},\ and\ \citenamefont
  {Fink}}]{egusa_multimaterial_2010}%
  \BibitemOpen
  \bibfield  {author} {\bibinfo {author} {\bibfnamefont {S.}~\bibnamefont
  {Egusa}}, \bibinfo {author} {\bibfnamefont {Z.}~\bibnamefont {Wang}},
  \bibinfo {author} {\bibfnamefont {N.}~\bibnamefont {Chocat}}, \bibinfo
  {author} {\bibfnamefont {Z.~M.}\ \bibnamefont {Ruff}}, \bibinfo {author}
  {\bibfnamefont {A.~M.}\ \bibnamefont {Stolyarov}}, \bibinfo {author}
  {\bibfnamefont {D.}~\bibnamefont {Shemuly}}, \bibinfo {author} {\bibfnamefont
  {F.}~\bibnamefont {Sorin}}, \bibinfo {author} {\bibfnamefont {P.~T.}\
  \bibnamefont {Rakich}}, \bibinfo {author} {\bibfnamefont {J.~D.}\
  \bibnamefont {Joannopoulos}}, \ and\ \bibinfo {author} {\bibfnamefont
  {Y.}~\bibnamefont {Fink}},\ }\bibfield  {title} {\enquote {\bibinfo {title}
  {Multimaterial piezoelectric fibres},}\ }\href {\doibase 10.1038/nmat2792}
  {\bibfield  {journal} {\bibinfo  {journal} {Nature Materials}\ }\textbf
  {\bibinfo {volume} {9}},\ \bibinfo {pages} {643--648} (\bibinfo {year}
  {2010})}\BibitemShut {NoStop}%
\bibitem [{\citenamefont {Gu}, \citenamefont {Gorgutsa},\ and\ \citenamefont
  {Skorobogatiy}(2010)}]{gu_soft_2010}%
  \BibitemOpen
  \bibfield  {author} {\bibinfo {author} {\bibfnamefont {J.~F.}\ \bibnamefont
  {Gu}}, \bibinfo {author} {\bibfnamefont {S.}~\bibnamefont {Gorgutsa}}, \ and\
  \bibinfo {author} {\bibfnamefont {M.}~\bibnamefont {Skorobogatiy}},\
  }\bibfield  {title} {\enquote {\bibinfo {title} {Soft capacitor fibers using
  conductive polymers for electronic textiles},}\ }\href@noop {} {\bibfield
  {journal} {\bibinfo  {journal} {Smart Materials and Structures}\ }\textbf
  {\bibinfo {volume} {19}},\ \bibinfo {pages} {115006} (\bibinfo {year}
  {2010})}\BibitemShut {NoStop}%
\bibitem [{\citenamefont {Khudiyev}\ \emph {et~al.}(2017)\citenamefont
  {Khudiyev}, \citenamefont {Clayton}, \citenamefont {Levy}, \citenamefont
  {Chocat}, \citenamefont {Gumennik}, \citenamefont {Stolyarov}, \citenamefont
  {Joannopoulos},\ and\ \citenamefont {Fink}}]{khudiyev_2017}%
  \BibitemOpen
  \bibfield  {author} {\bibinfo {author} {\bibfnamefont {T.}~\bibnamefont
  {Khudiyev}}, \bibinfo {author} {\bibfnamefont {J.}~\bibnamefont {Clayton}},
  \bibinfo {author} {\bibfnamefont {E.}~\bibnamefont {Levy}}, \bibinfo {author}
  {\bibfnamefont {N.}~\bibnamefont {Chocat}}, \bibinfo {author} {\bibfnamefont
  {A.}~\bibnamefont {Gumennik}}, \bibinfo {author} {\bibfnamefont {A.~M.}\
  \bibnamefont {Stolyarov}}, \bibinfo {author} {\bibfnamefont {J.}~\bibnamefont
  {Joannopoulos}}, \ and\ \bibinfo {author} {\bibfnamefont {Y.}~\bibnamefont
  {Fink}},\ }\bibfield  {title} {\enquote {\bibinfo {title} {Electrostrictive
  microelectromechanical fibres and textiles},}\ }\href@noop {} {\bibfield
  {journal} {\bibinfo  {journal} {Nature communications}\ }\textbf {\bibinfo
  {volume} {8}},\ \bibinfo {pages} {1435} (\bibinfo {year} {2017})}\BibitemShut
  {NoStop}%
\bibitem [{\citenamefont {Qu}\ \emph {et~al.}(2018)\citenamefont {Qu},
  \citenamefont {Nguyen-Dang}, \citenamefont {Page}, \citenamefont {Yan},
  \citenamefont {Das~Gupta}, \citenamefont {Rotaru}, \citenamefont {Rossi},
  \citenamefont {Favrod}, \citenamefont {Bartolomei},\ and\ \citenamefont
  {Sorin}}]{qu_superelastic_2018}%
  \BibitemOpen
  \bibfield  {author} {\bibinfo {author} {\bibfnamefont {Y.}~\bibnamefont
  {Qu}}, \bibinfo {author} {\bibfnamefont {T.}~\bibnamefont {Nguyen-Dang}},
  \bibinfo {author} {\bibfnamefont {A.~G.}\ \bibnamefont {Page}}, \bibinfo
  {author} {\bibfnamefont {W.}~\bibnamefont {Yan}}, \bibinfo {author}
  {\bibfnamefont {T.}~\bibnamefont {Das~Gupta}}, \bibinfo {author}
  {\bibfnamefont {G.~M.}\ \bibnamefont {Rotaru}}, \bibinfo {author}
  {\bibfnamefont {R.~M.}\ \bibnamefont {Rossi}}, \bibinfo {author}
  {\bibfnamefont {V.~D.}\ \bibnamefont {Favrod}}, \bibinfo {author}
  {\bibfnamefont {N.}~\bibnamefont {Bartolomei}}, \ and\ \bibinfo {author}
  {\bibfnamefont {F.}~\bibnamefont {Sorin}},\ }\bibfield  {title} {\enquote
  {\bibinfo {title} {Superelastic multimaterial electronic and photonic fibers
  and devices via thermal drawing},}\ }\href@noop {} {\bibfield  {journal}
  {\bibinfo  {journal} {Advanced Materials}\ }\textbf {\bibinfo {volume}
  {30}},\ \bibinfo {pages} {1707251} (\bibinfo {year} {2018})}\BibitemShut
  {NoStop}%
\bibitem [{\citenamefont {Stolyarov}\ \emph {et~al.}(2012)\citenamefont
  {Stolyarov}, \citenamefont {Wei}, \citenamefont {Sorin}, \citenamefont
  {Lestoquoy}, \citenamefont {Joannopoulos},\ and\ \citenamefont
  {Fink}}]{stolyarov_2012}%
  \BibitemOpen
  \bibfield  {author} {\bibinfo {author} {\bibfnamefont {A.~M.}\ \bibnamefont
  {Stolyarov}}, \bibinfo {author} {\bibfnamefont {L.}~\bibnamefont {Wei}},
  \bibinfo {author} {\bibfnamefont {F.}~\bibnamefont {Sorin}}, \bibinfo
  {author} {\bibfnamefont {G.}~\bibnamefont {Lestoquoy}}, \bibinfo {author}
  {\bibfnamefont {J.~D.}\ \bibnamefont {Joannopoulos}}, \ and\ \bibinfo
  {author} {\bibfnamefont {Y.}~\bibnamefont {Fink}},\ }\bibfield  {title}
  {\enquote {\bibinfo {title} {Fabrication and characterization of fibers with
  built-in liquid crystal channels and electrodes for transverse incident-light
  modulation},}\ }\href@noop {} {\bibfield  {journal} {\bibinfo  {journal}
  {Applied Physics Letters}\ }\textbf {\bibinfo {volume} {101}},\ \bibinfo
  {pages} {011108} (\bibinfo {year} {2012})}\BibitemShut {NoStop}%
\bibitem [{\citenamefont {Yan}\ \emph {et~al.}(2017)\citenamefont {Yan},
  \citenamefont {Qu}, \citenamefont {Gupta}, \citenamefont {Darga},
  \citenamefont {Nguy{\^e}n}, \citenamefont {Page}, \citenamefont {Rossi},
  \citenamefont {Ceriotti},\ and\ \citenamefont {Sorin}}]{yan_2017}%
  \BibitemOpen
  \bibfield  {author} {\bibinfo {author} {\bibfnamefont {W.}~\bibnamefont
  {Yan}}, \bibinfo {author} {\bibfnamefont {Y.}~\bibnamefont {Qu}}, \bibinfo
  {author} {\bibfnamefont {T.~D.}\ \bibnamefont {Gupta}}, \bibinfo {author}
  {\bibfnamefont {A.}~\bibnamefont {Darga}}, \bibinfo {author} {\bibfnamefont
  {D.~T.}\ \bibnamefont {Nguy{\^e}n}}, \bibinfo {author} {\bibfnamefont
  {A.~G.}\ \bibnamefont {Page}}, \bibinfo {author} {\bibfnamefont
  {M.}~\bibnamefont {Rossi}}, \bibinfo {author} {\bibfnamefont
  {M.}~\bibnamefont {Ceriotti}}, \ and\ \bibinfo {author} {\bibfnamefont
  {F.}~\bibnamefont {Sorin}},\ }\bibfield  {title} {\enquote {\bibinfo {title}
  {Semiconducting nanowire-based optoelectronic fibers},}\ }\href@noop {}
  {\bibfield  {journal} {\bibinfo  {journal} {Advanced Materials}\ }\textbf
  {\bibinfo {volume} {29}},\ \bibinfo {pages} {1700681} (\bibinfo {year}
  {2017})}\BibitemShut {NoStop}%
\bibitem [{\citenamefont {Nguyen-Dang}\ \emph
  {et~al.}(2017{\natexlab{a}})\citenamefont {Nguyen-Dang}, \citenamefont
  {de~Luca}, \citenamefont {Yan}, \citenamefont {Qu}, \citenamefont {Page},
  \citenamefont {Volpi}, \citenamefont {Das~Gupta}, \citenamefont {Lacour},\
  and\ \citenamefont {Sorin}}]{nguyen-dang_controlled_2017}%
  \BibitemOpen
  \bibfield  {author} {\bibinfo {author} {\bibfnamefont {T.}~\bibnamefont
  {Nguyen-Dang}}, \bibinfo {author} {\bibfnamefont {A.~C.}\ \bibnamefont
  {de~Luca}}, \bibinfo {author} {\bibfnamefont {W.}~\bibnamefont {Yan}},
  \bibinfo {author} {\bibfnamefont {Y.}~\bibnamefont {Qu}}, \bibinfo {author}
  {\bibfnamefont {A.~G.}\ \bibnamefont {Page}}, \bibinfo {author}
  {\bibfnamefont {M.}~\bibnamefont {Volpi}}, \bibinfo {author} {\bibfnamefont
  {T.}~\bibnamefont {Das~Gupta}}, \bibinfo {author} {\bibfnamefont {S.~P.}\
  \bibnamefont {Lacour}}, \ and\ \bibinfo {author} {\bibfnamefont
  {F.}~\bibnamefont {Sorin}},\ }\bibfield  {title} {\enquote {\bibinfo {title}
  {Controlled {Sub}-{Micrometer} {Hierarchical} {Textures} {Engineered} in
  {Polymeric} {Fibers} and {Microchannels} via {Thermal} {Drawing}},}\ }\href
  {\doibase 10.1002/adfm.201605935} {\bibfield  {journal} {\bibinfo  {journal}
  {Advanced Functional Materials}\ }\textbf {\bibinfo {volume} {27}},\ \bibinfo
  {pages} {1605935} (\bibinfo {year} {2017}{\natexlab{a}})}\BibitemShut
  {NoStop}%
\bibitem [{\citenamefont {Guo}\ \emph {et~al.}(2017)\citenamefont {Guo},
  \citenamefont {Jiang}, \citenamefont {Grena}, \citenamefont {Kimbrough},
  \citenamefont {Thompson}, \citenamefont {Fink}, \citenamefont {Sontheimer},
  \citenamefont {Yoshinobu},\ and\ \citenamefont {Jia}}]{guo_polymer_2017}%
  \BibitemOpen
  \bibfield  {author} {\bibinfo {author} {\bibfnamefont {Y.}~\bibnamefont
  {Guo}}, \bibinfo {author} {\bibfnamefont {S.}~\bibnamefont {Jiang}}, \bibinfo
  {author} {\bibfnamefont {B.~J.~B.}\ \bibnamefont {Grena}}, \bibinfo {author}
  {\bibfnamefont {I.~F.}\ \bibnamefont {Kimbrough}}, \bibinfo {author}
  {\bibfnamefont {E.~G.}\ \bibnamefont {Thompson}}, \bibinfo {author}
  {\bibfnamefont {Y.}~\bibnamefont {Fink}}, \bibinfo {author} {\bibfnamefont
  {H.}~\bibnamefont {Sontheimer}}, \bibinfo {author} {\bibfnamefont
  {T.}~\bibnamefont {Yoshinobu}}, \ and\ \bibinfo {author} {\bibfnamefont
  {X.}~\bibnamefont {Jia}},\ }\bibfield  {title} {\enquote {\bibinfo {title}
  {Polymer {Composite} with {Carbon} {Nanofibers} {Aligned} during {Thermal}
  {Drawing} as a {Microelectrode} for {Chronic} {Neural} {Interfaces}},}\
  }\href {\doibase 10.1021/acsnano.6b07550} {\bibfield  {journal} {\bibinfo
  {journal} {ACS Nano}\ }\textbf {\bibinfo {volume} {11}},\ \bibinfo {pages}
  {6574--6585} (\bibinfo {year} {2017})}\BibitemShut {NoStop}%
\bibitem [{\citenamefont {Chen}, \citenamefont {Canales},\ and\ \citenamefont
  {Anikeeva}(2017)}]{chen_2017}%
  \BibitemOpen
  \bibfield  {author} {\bibinfo {author} {\bibfnamefont {R.}~\bibnamefont
  {Chen}}, \bibinfo {author} {\bibfnamefont {A.}~\bibnamefont {Canales}}, \
  and\ \bibinfo {author} {\bibfnamefont {P.}~\bibnamefont {Anikeeva}},\
  }\bibfield  {title} {\enquote {\bibinfo {title} {Neural recording and
  modulation technologies},}\ }\href@noop {} {\bibfield  {journal} {\bibinfo
  {journal} {Nature Reviews Materials}\ }\textbf {\bibinfo {volume} {2}},\
  \bibinfo {pages} {16093} (\bibinfo {year} {2017})}\BibitemShut {NoStop}%
\bibitem [{\citenamefont {Pone}\ \emph {et~al.}(2006)\citenamefont {Pone},
  \citenamefont {Dubois}, \citenamefont {Guo}, \citenamefont {Gao},
  \citenamefont {Dupuis}, \citenamefont {Boismenu}, \citenamefont {Lacroix},\
  and\ \citenamefont {Skorobogatiy}}]{pone_2006}%
  \BibitemOpen
  \bibfield  {author} {\bibinfo {author} {\bibfnamefont {E.}~\bibnamefont
  {Pone}}, \bibinfo {author} {\bibfnamefont {C.}~\bibnamefont {Dubois}},
  \bibinfo {author} {\bibfnamefont {N.}~\bibnamefont {Guo}}, \bibinfo {author}
  {\bibfnamefont {Y.}~\bibnamefont {Gao}}, \bibinfo {author} {\bibfnamefont
  {A.}~\bibnamefont {Dupuis}}, \bibinfo {author} {\bibfnamefont
  {F.}~\bibnamefont {Boismenu}}, \bibinfo {author} {\bibfnamefont
  {S.}~\bibnamefont {Lacroix}}, \ and\ \bibinfo {author} {\bibfnamefont
  {M.}~\bibnamefont {Skorobogatiy}},\ }\bibfield  {title} {\enquote {\bibinfo
  {title} {Drawing of the hollow all-polymer {Bragg} fibers},}\ }\href@noop {}
  {\bibfield  {journal} {\bibinfo  {journal} {Optics Express}\ }\textbf
  {\bibinfo {volume} {14}},\ \bibinfo {pages} {5838--5852} (\bibinfo {year}
  {2006})}\BibitemShut {NoStop}%
\bibitem [{\citenamefont {Xue}\ \emph {et~al.}(2005)\citenamefont {Xue},
  \citenamefont {Tanner}, \citenamefont {Barton}, \citenamefont {Lwin},
  \citenamefont {Large},\ and\ \citenamefont
  {Poladian}}]{xue_fabrication_2005}%
  \BibitemOpen
  \bibfield  {author} {\bibinfo {author} {\bibfnamefont {S.~C.}\ \bibnamefont
  {Xue}}, \bibinfo {author} {\bibfnamefont {R.~I.}\ \bibnamefont {Tanner}},
  \bibinfo {author} {\bibfnamefont {G.~W.}\ \bibnamefont {Barton}}, \bibinfo
  {author} {\bibfnamefont {R.}~\bibnamefont {Lwin}}, \bibinfo {author}
  {\bibfnamefont {M.~C.~J.}\ \bibnamefont {Large}}, \ and\ \bibinfo {author}
  {\bibfnamefont {L.}~\bibnamefont {Poladian}},\ }\bibfield  {title} {\enquote
  {\bibinfo {title} {Fabrication of {Microstructured} {Optical} {Fibers}-{Part}
  {I}: {Problem} {Formulation} and {Numerical} {Modeling} of {Transient} {Draw}
  {Process}},}\ }\href
  {http://www.osapublishing.org/abstract.cfm?uri=jlt-23-7-2245} {\bibfield
  {journal} {\bibinfo  {journal} {Journal of Lightwave Technology}\ }\textbf
  {\bibinfo {volume} {23}},\ \bibinfo {pages} {2245} (\bibinfo {year}
  {2005})}\BibitemShut {NoStop}%
\bibitem [{\citenamefont {Xue}\ \emph {et~al.}(2017)\citenamefont {Xue},
  \citenamefont {Barton}, \citenamefont {Fleming},\ and\ \citenamefont
  {Argyros}}]{xue_heat_2017}%
  \BibitemOpen
  \bibfield  {author} {\bibinfo {author} {\bibfnamefont {S.}~\bibnamefont
  {Xue}}, \bibinfo {author} {\bibfnamefont {G.}~\bibnamefont {Barton}},
  \bibinfo {author} {\bibfnamefont {S.}~\bibnamefont {Fleming}}, \ and\
  \bibinfo {author} {\bibfnamefont {A.}~\bibnamefont {Argyros}},\ }\bibfield
  {title} {\enquote {\bibinfo {title} {Heat {Transfer} {Modeling} of the
  {Capillary} {Fiber} {Drawing} {Process}},}\ }\href {\doibase
  10.1115/1.4035714} {\bibfield  {journal} {\bibinfo  {journal} {Journal of
  Heat Transfer}\ }\textbf {\bibinfo {volume} {139}},\ \bibinfo {pages}
  {072001} (\bibinfo {year} {2017})}\BibitemShut {NoStop}%
\bibitem [{\citenamefont {Nguyen-Dang}\ \emph
  {et~al.}(2017{\natexlab{b}})\citenamefont {Nguyen-Dang}, \citenamefont
  {Page}, \citenamefont {Qu}, \citenamefont {Volpi}, \citenamefont {Yan},\ and\
  \citenamefont {Sorin}}]{nguyen-dang_multi-material_2017}%
  \BibitemOpen
  \bibfield  {author} {\bibinfo {author} {\bibfnamefont {T.}~\bibnamefont
  {Nguyen-Dang}}, \bibinfo {author} {\bibfnamefont {A.~G.}\ \bibnamefont
  {Page}}, \bibinfo {author} {\bibfnamefont {Y.}~\bibnamefont {Qu}}, \bibinfo
  {author} {\bibfnamefont {M.}~\bibnamefont {Volpi}}, \bibinfo {author}
  {\bibfnamefont {W.}~\bibnamefont {Yan}}, \ and\ \bibinfo {author}
  {\bibfnamefont {F.}~\bibnamefont {Sorin}},\ }\bibfield  {title} {\enquote
  {\bibinfo {title} {Multi-material micro-electromechanical fibers with
  bendable functional domains},}\ }\href
  {http://stacks.iop.org/0022-3727/50/i=14/a=144001} {\bibfield  {journal}
  {\bibinfo  {journal} {Journal of Physics D: Applied Physics}\ }\textbf
  {\bibinfo {volume} {50}},\ \bibinfo {pages} {144001} (\bibinfo {year}
  {2017}{\natexlab{b}})}\BibitemShut {NoStop}%
\bibitem [{\citenamefont {Matovich}\ and\ \citenamefont
  {Pearson}(1969)}]{matovich_1969}%
  \BibitemOpen
  \bibfield  {author} {\bibinfo {author} {\bibfnamefont {M.~A.}\ \bibnamefont
  {Matovich}}\ and\ \bibinfo {author} {\bibfnamefont {J.~R.}\ \bibnamefont
  {Pearson}},\ }\bibfield  {title} {\enquote {\bibinfo {title} {{Spinning a
  molten threadline: Steady-State Isothermal Viscous Flows}},}\ }\href
  {\doibase 10.1021/i160031a023} {\bibfield  {journal} {\bibinfo  {journal}
  {Industrial and Engineering Chemistry Fundamentals}\ }\textbf {\bibinfo
  {volume} {8}},\ \bibinfo {pages} {512--520} (\bibinfo {year}
  {1969})}\BibitemShut {NoStop}%
\bibitem [{\citenamefont {Shah}\ and\ \citenamefont
  {Pearson}(1972)}]{shah_1972}%
  \BibitemOpen
  \bibfield  {author} {\bibinfo {author} {\bibfnamefont {Y.~T.}\ \bibnamefont
  {Shah}}\ and\ \bibinfo {author} {\bibfnamefont {J.~R.~A.}\ \bibnamefont
  {Pearson}},\ }\bibfield  {title} {\enquote {\bibinfo {title} {On the
  stability of nonisothermal fiber spinning},}\ }\href@noop {} {\bibfield
  {journal} {\bibinfo  {journal} {Industrial \& Engineering Chemistry
  Fundamentals}\ }\textbf {\bibinfo {volume} {11}},\ \bibinfo {pages}
  {145--149} (\bibinfo {year} {1972})}\BibitemShut {NoStop}%
\bibitem [{\citenamefont {German}\ and\ \citenamefont
  {Khayat}(2008)}]{german_2008}%
  \BibitemOpen
  \bibfield  {author} {\bibinfo {author} {\bibfnamefont {R.}~\bibnamefont
  {German}}\ and\ \bibinfo {author} {\bibfnamefont {R.~E.}\ \bibnamefont
  {Khayat}},\ }\bibfield  {title} {\enquote {\bibinfo {title} {{Interplay
  Between Inertia and Elasticity in Film Casting}},}\ }\href {\doibase
  10.1115/1.2956592} {\bibfield  {journal} {\bibinfo  {journal} {Journal of
  Fluids Engineering}\ }\textbf {\bibinfo {volume} {130}},\ \bibinfo {pages}
  {081501} (\bibinfo {year} {2008})}\BibitemShut {NoStop}%
\bibitem [{\citenamefont {Bechert}\ and\ \citenamefont
  {Scheid}(2017)}]{bechert_combined_2017}%
  \BibitemOpen
  \bibfield  {author} {\bibinfo {author} {\bibfnamefont {M.}~\bibnamefont
  {Bechert}}\ and\ \bibinfo {author} {\bibfnamefont {B.}~\bibnamefont
  {Scheid}},\ }\bibfield  {title} {\enquote {\bibinfo {title} {Combined
  influence of inertia, gravity, and surface tension on the linear stability of
  {Newtonian} fiber spinning},}\ }\href {\doibase
  10.1103/PhysRevFluids.2.113905} {\bibfield  {journal} {\bibinfo  {journal}
  {Physical Review Fluids}\ }\textbf {\bibinfo {volume} {2}},\ \bibinfo {pages}
  {113905} (\bibinfo {year} {2017})}\BibitemShut {NoStop}%
\bibitem [{\citenamefont {Schultz}\ and\ \citenamefont
  {Davis}(1982)}]{schultz_1982}%
  \BibitemOpen
  \bibfield  {author} {\bibinfo {author} {\bibfnamefont {W.~W.}\ \bibnamefont
  {Schultz}}\ and\ \bibinfo {author} {\bibfnamefont {S.~H.}\ \bibnamefont
  {Davis}},\ }\bibfield  {title} {\enquote {\bibinfo {title} {One-{Dimensional}
  {Liquid} {Fibers}},}\ }\href@noop {} {\bibfield  {journal} {\bibinfo
  {journal} {Journal of Rheology (1978-present)}\ }\textbf {\bibinfo {volume}
  {26}},\ \bibinfo {pages} {331--345} (\bibinfo {year} {1982})}\BibitemShut
  {NoStop}%
\bibitem [{\citenamefont {Scheid}\ \emph {et~al.}(2009)\citenamefont {Scheid},
  \citenamefont {Quiligotti}, \citenamefont {Tran}, \citenamefont {Gy},\ and\
  \citenamefont {Stone}}]{scheid_stabilization_2009}%
  \BibitemOpen
  \bibfield  {author} {\bibinfo {author} {\bibfnamefont {B.}~\bibnamefont
  {Scheid}}, \bibinfo {author} {\bibfnamefont {S.}~\bibnamefont {Quiligotti}},
  \bibinfo {author} {\bibfnamefont {B.}~\bibnamefont {Tran}}, \bibinfo {author}
  {\bibfnamefont {R.}~\bibnamefont {Gy}}, \ and\ \bibinfo {author}
  {\bibfnamefont {H.~A.}\ \bibnamefont {Stone}},\ }\bibfield  {title} {\enquote
  {\bibinfo {title} {On the (de)stabilization of draw resonance due to
  cooling},}\ }\href {\doibase 10.1017/S0022112009007836} {\bibfield  {journal}
  {\bibinfo  {journal} {Journal of Fluid Mechanics}\ }\textbf {\bibinfo
  {volume} {636}},\ \bibinfo {pages} {155--176} (\bibinfo {year}
  {2009})}\BibitemShut {NoStop}%
\bibitem [{\citenamefont {Scheid}, \citenamefont {van Nierop},\ and\
  \citenamefont {Stone}(2012)}]{scheid_2012}%
  \BibitemOpen
  \bibfield  {author} {\bibinfo {author} {\bibfnamefont {B.}~\bibnamefont
  {Scheid}}, \bibinfo {author} {\bibfnamefont {E.~A.}\ \bibnamefont {van
  Nierop}}, \ and\ \bibinfo {author} {\bibfnamefont {H.~A.}\ \bibnamefont
  {Stone}},\ }\bibfield  {title} {\enquote {\bibinfo {title}
  {Thermocapillary-assisted pulling of contact-free liquid films},}\ }\href
  {\doibase 10.1063/1.3692097} {\bibfield  {journal} {\bibinfo  {journal}
  {Physics of Fluids}\ }\textbf {\bibinfo {volume} {24}},\ \bibinfo {pages}
  {032107} (\bibinfo {year} {2012})}\BibitemShut {NoStop}%
\bibitem [{\citenamefont {Garvey}\ \emph {et~al.}(1996)\citenamefont {Garvey},
  \citenamefont {Zimmerman}, \citenamefont {Young}, \citenamefont {Tostenrude},
  \citenamefont {Townsend}, \citenamefont {Zhou}, \citenamefont {Lobel},
  \citenamefont {Dayton}, \citenamefont {Wittorf}, \citenamefont {Kuzyk},
  \citenamefont {Sounick},\ and\ \citenamefont
  {Dirk}}]{garvey_single-mode_1996}%
  \BibitemOpen
  \bibfield  {author} {\bibinfo {author} {\bibfnamefont {D.~W.}\ \bibnamefont
  {Garvey}}, \bibinfo {author} {\bibfnamefont {K.}~\bibnamefont {Zimmerman}},
  \bibinfo {author} {\bibfnamefont {P.}~\bibnamefont {Young}}, \bibinfo
  {author} {\bibfnamefont {J.}~\bibnamefont {Tostenrude}}, \bibinfo {author}
  {\bibfnamefont {J.~S.}\ \bibnamefont {Townsend}}, \bibinfo {author}
  {\bibfnamefont {Z.}~\bibnamefont {Zhou}}, \bibinfo {author} {\bibfnamefont
  {M.}~\bibnamefont {Lobel}}, \bibinfo {author} {\bibfnamefont
  {M.}~\bibnamefont {Dayton}}, \bibinfo {author} {\bibfnamefont
  {R.}~\bibnamefont {Wittorf}}, \bibinfo {author} {\bibfnamefont {M.~G.}\
  \bibnamefont {Kuzyk}}, \bibinfo {author} {\bibfnamefont {J.}~\bibnamefont
  {Sounick}}, \ and\ \bibinfo {author} {\bibfnamefont {C.~W.}\ \bibnamefont
  {Dirk}},\ }\bibfield  {title} {\enquote {\bibinfo {title} {Single-mode
  nonlinear-optical polymer fibers},}\ }\href {\doibase
  10.1364/JOSAB.13.002017} {\bibfield  {journal} {\bibinfo  {journal} {Journal
  of the Optical Society of America B}\ }\textbf {\bibinfo {volume} {13}},\
  \bibinfo {pages} {2017} (\bibinfo {year} {1996})}\BibitemShut {NoStop}%
\bibitem [{\citenamefont {Doedel}\ \emph {et~al.}()\citenamefont {Doedel},
  \citenamefont {Champneys}, \citenamefont {Fairgrieve}, \citenamefont
  {Kuznetsov}, \citenamefont {Sandstede},\ and\ \citenamefont {Wang}}]{auto}%
  \BibitemOpen
  \bibfield  {author} {\bibinfo {author} {\bibfnamefont {E.~J.}\ \bibnamefont
  {Doedel}}, \bibinfo {author} {\bibfnamefont {A.~R.}\ \bibnamefont
  {Champneys}}, \bibinfo {author} {\bibfnamefont {T.~F.}\ \bibnamefont
  {Fairgrieve}}, \bibinfo {author} {\bibfnamefont {Y.~A.}\ \bibnamefont
  {Kuznetsov}}, \bibinfo {author} {\bibfnamefont {B.}~\bibnamefont
  {Sandstede}}, \ and\ \bibinfo {author} {\bibfnamefont {X.}~\bibnamefont
  {Wang}},\ }\href@noop {} {\enquote {\bibinfo {title} {{AUTO} 97:
  {C}ontinuation and {B}ifurcation {S}oftware {F}or {O}rdinary {D}ifferential
  {E}quations (with {H}om{C}ont)},}\ }\BibitemShut {NoStop}%
\bibitem [{\citenamefont {Saphiannikova}\ \emph {et~al.}(2012)\citenamefont
  {Saphiannikova}, \citenamefont {Skipa}, \citenamefont {Lellinger},
  \citenamefont {Alig},\ and\ \citenamefont
  {Heinrich}}]{saphiannikova_superposition_2012}%
  \BibitemOpen
  \bibfield  {author} {\bibinfo {author} {\bibfnamefont {M.}~\bibnamefont
  {Saphiannikova}}, \bibinfo {author} {\bibfnamefont {T.}~\bibnamefont
  {Skipa}}, \bibinfo {author} {\bibfnamefont {D.}~\bibnamefont {Lellinger}},
  \bibinfo {author} {\bibfnamefont {I.}~\bibnamefont {Alig}}, \ and\ \bibinfo
  {author} {\bibfnamefont {G.}~\bibnamefont {Heinrich}},\ }\bibfield  {title}
  {\enquote {\bibinfo {title} {Superposition approach for description of
  electrical conductivity in sheared {MWNT}/polycarbonate melts},}\ }\href@noop
  {} {\bibfield  {journal} {\bibinfo  {journal} {Express Polymer Letters}\
  }\textbf {\bibinfo {volume} {6}},\ \bibinfo {pages} {438--453} (\bibinfo
  {year} {2012})}\BibitemShut {NoStop}%
\end{thebibliography}
\end{document}


\newcommand{\hl}[1]{\textcolor{red}{#1}}
\newcommand{\comM}[1]{\textcolor{cyan}{#1}}
\def\theequation{S\arabic{equation}}
\newcommand{\beginsupplement}{%
        \setcounter{table}{0}
        \renewcommand{\thetable}{S\arabic{table}}%
        \setcounter{figure}{0}
        \renewcommand{\thefigure}{S\arabic{figure}}%
     }

\title{Unraveling radial dependency effects in fiber thermal drawing\\Supplementary material}

\author{Alexis G. Page}
\author{Mathias Bechert}
\author{Fran\c{c}ois Gallaire}
\author{Fabien Sorin}

\date{\today}

\maketitle
\beginsupplement

\section{Model derivation}
\subsection{Governing equations}
We follow the derivation as presented by \citet{bechert_combined_2017} for fibers and extend the model equations to cover thermal effects, analogue to \citet{scheid_stabilization_2009}. Moreover, we analyze higher order terms of the velocity field.\par

We model the fiber material as an incompressible, Newtonian fluid and assume negligible effects of inertia, gravity and surface tension ($Re<10^{-6}$, $Re/Fr<1$, $Ca>10^6$). We further neglect all time derivatives as we are interested in steady solutions only. The corresponding governing equations are the continuity, momentum and heat equations, which can be written as, respectively,
\begin{subequations}
\label{eq:governing}
\begin{align}
\label{eq:gov_cont}
\boldsymbol\nabla\cdot\boldsymbol u &= 0,\\
\label{eq:gov_mom}
\boldsymbol\nabla\cdot\boldsymbol\sigma &= 0,\\
\rho\,c_p\,\partial_z T &= k\,\nabla^2\,T,
\end{align}
\end{subequations}
with $\boldsymbol\nabla = \boldsymbol e_r\partial_r + \boldsymbol e_\phi\partial_{\phi}/r+\boldsymbol e_z\partial_z$, velocity $\boldsymbol u = (u,v)$, $u$ and $v$ denoting the radial and axial velocity components, and stress tensor $\boldsymbol\sigma = -p\,\boldsymbol{1} + \eta\,\left(\boldsymbol\nabla\boldsymbol u + \left(\boldsymbol\nabla\boldsymbol u\right)^T\right)$, with pressure $p$ and viscosity~$\eta$. $\rho$, $c_p$ and $k$ are the material density, heat capacity and heat conductivity, respectively, which are all assumed to be constant.\\
The equations are then made dimensionless applying the following transformation rule,
\begin{align}
\begin{split}
r\rightarrow\alpha\,L\,r,\,\,\,\,z\rightarrow L\,z,\,\,\,\,u\rightarrow \alpha\,v_0\,u,\,\,\,\,v\rightarrow v_0,\,\,\,\,\eta\rightarrow\eta_{\rm min}\,\eta,\\\,\,\,\,p\rightarrow\eta_{\rm min} \frac{v_0}{L}\,p,\,\,\,\,\sigma\rightarrow\eta_{\rm min} \frac{v_0}{L}\,\sigma,\,\,\,\,T\rightarrow T_{\rm max}\,T,
\end{split}
\end{align}
where the subscript `$0$' denotes the position at $z=0$, $T_{\rm max}$ the maximum temperature of the surrounding air (see below) and $\eta_{\rm min}$ the corresponding viscosity. Note that even though the material does not necessarily reach $T_{\rm max}$ and $\eta_{\rm min}$, it still approaches these values closely during the drawing, which makes them well-suited for scaling. $\alpha=R_0/L$, the ratio of preform radius to drawing length, is also referred to as the fiber parameter.\par

At the fiber surface $r = R(z)$, we impose the following boundary conditions,
\begin{subequations}
\begin{align}
v|_{R}\partial_zR - u|_R &= 0,\\
\boldsymbol n\cdot\boldsymbol\sigma |_R &= 0,\\
\label{eq:NBC}
-k\,\boldsymbol n\cdot\boldsymbol\nabla T|_R &= h\,\left(T|_{R}-T_{\rm air}(z)\right),
\end{align}
\end{subequations}
which are also often referred to as the kinematic, free stress and Newton's cooling boundary conditions, respectively. The vector normal to the fiber surface is denoted by $\boldsymbol n$, $h$ denotes the heat transfer coefficient between the fiber material and the surrounding air, assumed to be constant, and $T_{\rm air}$ the temperature profile of the surrounding air. While radiation is neglected in (\ref{eq:NBC}), we adapt $h$ and $T_{\rm air}$ later by fitting the radius profile of the fiber (see below), which means we are using an effective heat transfer coefficient and air temperature profile to cover both convection and radiation using the simple form of (\ref{eq:NBC}). Note that it is also possible to derive Eq.~(\ref{eq:NBC}) by linearizing the radiative heat transfer \cite{scheid_2012}.\par

We assume a centered Gaussian profile for the air temperature in the furnace along the drawing axis, which reads in dimensionless form
\begin{align}
\label{eq:T_prof}
T_{\rm air} = \frac{\Lambda\,\exp\left[-\frac{\left(z-0.5\right)^2}{\Delta^2}\right]+1}{\Lambda+1},
\end{align}
and is determined by the dimensionless amplitude $\Lambda$ and width $\Delta$. The domain of modeling, i.e., the drawing length $L$, is set to $40\,\rm{cm}$ around the maximum of the ambient temperature $T_{\rm max}$, which is also used for scaling. With this choice of domain, the temperature at the boundaries is small enough so that the resulting high viscosities lead to negligible fiber deformation.

\subsection{Cross-section averaged model}
The variables are expanded in orders of $\alpha^2$, i.e.,
\begin{align}
\boldsymbol x = \boldsymbol x^{(0)} + \alpha^2 \boldsymbol x^{(1)} + \mathcal{O}\left(\alpha^4\right),
\end{align}
with state vector $\boldsymbol x = (p,\boldsymbol u,\eta,T)$. Analogue to the way \citet{scheid_stabilization_2009} included non-isothermal effects in one-dimensional film casting, we assume a parabolic temperature field in $r$ with averaged temperature $\langle T \rangle$, which fulfills boundary condition (\ref{eq:NBC}),
\begin{align}
\label{eq:T_quadr}
T^{(0)} = \langle T^{(0)}\rangle + Bi\, \frac{\left(\langle T^{(0)}\rangle-T_{\rm air}\right) R}{4+Bi\,R}\left(1-\frac{r^2}{R^2}\right),
\end{align}
with Biot number $Bi = R_0\, h/k$. This enables us to cross-section average the leading order of the governing equations~(\ref{eq:governing}) by integration to obtain
\begin{subequations}
\label{eq:cont_mom_heat}
\begin{align}
\partial_z \left(R^2\,v^{(0)}\right) &= 0,\\
\label{eq:mom}
\partial_z \left(R^2\,\langle\eta^{(0)}\rangle\,\partial_z v^{(0)}\right) &= 0,\\
\label{eq:heat}
v^{(0)}\partial_z \langle T^{(0)}\rangle &= -St\,\frac{\langle T^{(0)}\rangle-T_{\rm air}}{R(1+Bi\,R/4)},
\end{align}
\end{subequations}
with Stanton number $St = 2\,Bi/(\alpha Pe)$, where the Peclet number is given by $Pe=\rho c_p v_0 R_0 / k$. This derivation also yields $v^{(0)} = v^{(0)}(z)$, i.e., the axial velocity at leading order does not vary along the radius.\par
Equations~(\ref{eq:cont_mom_heat}) are supplemented by Arrhenius' law, which links the heat equation to the rest of the system and is given by
\begin{align}
\label{eq:arrhenius}
\eta = \exp\left[\mu\left(\frac{1}{T}-1\right)\right].
\end{align}
The dimensionless activation energy $\mu = E_a / (\bar R\,T_{\rm max})$, with dimensional activation energy $E_a$ and Gas constant $\bar R$, quantifies the sensitivity of the viscosity on temperature changes. The boundary conditions,
\begin{subequations}
\label{eq:BC}
\begin{align}
v^{(0)}(0) = R(0) &= 1,\\
\langle T^{(0)}\rangle(0) &= T_{\rm air}(0),\\
v^{(0)} (1) &= Dr^2,
\end{align}
\end{subequations}
close the cross-section averaged model.

\subsection{Higher order terms}
We now investigate the $\alpha^2$-order of the momentum equation~(\ref{eq:gov_mom}), which yields
\begin{align}
\label{eq:v1}
\partial_r\left[\eta^{(0)}r\left(\partial_rv^{(1)} - \frac{r}{2}\partial_{zz}v^{(0)}\right)\right] + 3\,r\,\partial_z\left(\eta^{(0)}\partial_zv^{(0)}\right) = 0.
\end{align}
To solve this equation for $v^{(1)}$ by numerical integration, we require the condition
\begin{align}
\label{eq:v1BC}
\langle v^{(1)}\rangle = 0,
\end{align}
so as to preserve the fact that at leading order, $v^{(0)}$ is equal to the averaged velocity $\langle v\rangle$. The $\alpha^2$-order of the continuity equation~(\ref{eq:gov_cont}) then leads to
\begin{align}
u^{(1)} = -\frac{1}{r}\int_0^r dr\,r\,\partial_z v^{(1)}.
\end{align}
While we introduced a radial dependent temperature $T^{(0)}$ with Eq.~(\ref{eq:T_quadr}) in the derivation of the leading order model, and consequently a radial dependent viscosity $\eta^{(0)}$ as well, the final cross-section averaged model widely used in literature uses merely the averaged variables $\langle T^{(0)}\rangle$ and $\langle \eta^{(0)}\rangle$. In our model we use the velocity field up to $\alpha^2$-order as well as $T^{(0)}$ and $\eta^{(0)}$ in order to investigate the radial distribution of fiber properties.

\subsection{Fiber conductivity}
\begin{figure}
\centering
\includegraphics[width=0.9\textwidth]{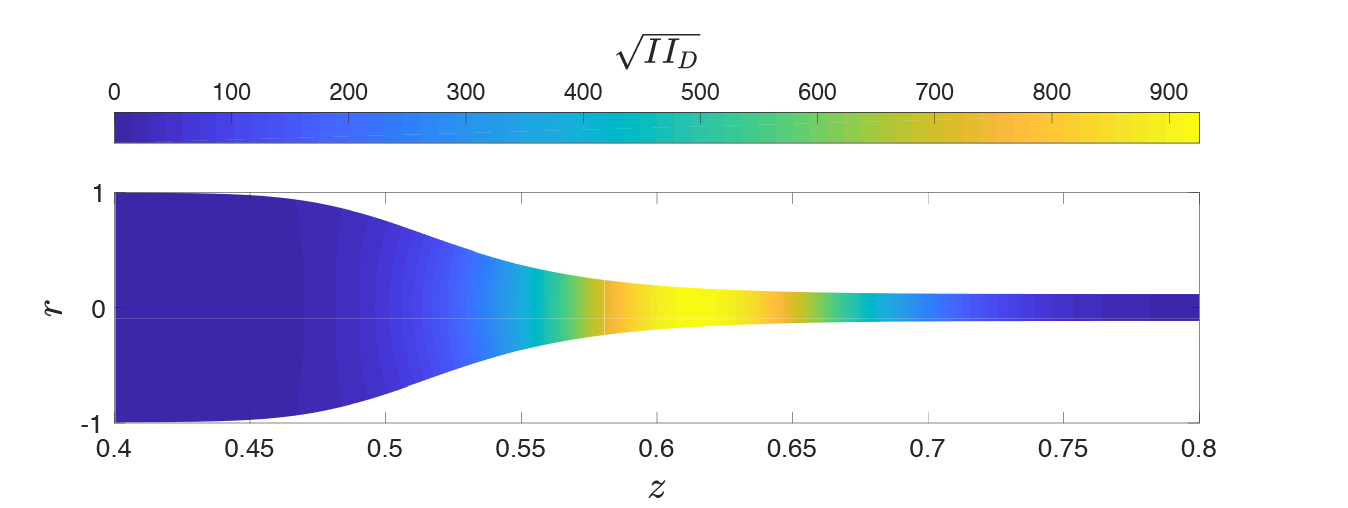}
\caption{$\sqrt{II_D}$-field for a PSu fiber drawn at $Dr=8.31$.}
\label{fig:IID}
\end{figure}
Inspired by the model presented by \citet{saphiannikova_superposition_2012}, we extend their kinetic equation for the effective filler concentration $p_e$ to use the second invariant of the deformation tensor $\sqrt{II_D}$, defined in its dimensionless form as
\begin{align}
\sqrt{II_D} = \sqrt{\left(\alpha\,\partial_zu+\partial_r v/\alpha\right)^2+2\left(u/r\right)^2+2\left(\partial_ru\right)^2+2\left(\partial_zv\right)^2},
\end{align}
instead of the shear rate, which then yields
\begin{align}
\label{eq:kinetic}
\partial_t p_e = -a\,\eta\,\sqrt{II_D}^{\,m}(p_e-p_c).
\end{align}
Figure~\ref{fig:IID} shows the distribution of $\sqrt{II_D}$ in a PSu fiber. The dimensionless destruction rate $a$ is used as a free fit parameter and is proportional to $(v_0/L)^{\,m-1}\,\eta_{\rm min}$, while $m$ is set to $1.5$, following \citet{saphiannikova_superposition_2012}. The critical concentration $p_c$, below which no conductive pathway exists can be eliminated later.\par
Integrating Eq.~(\ref{eq:kinetic}) yields
\begin{align}
\label{eq:pe_sol}
p_e = p_c + (p_{\rm in} - p_c)\,\exp\left(-a\int_0^tdt\,\sqrt{II_D}^{\,m}\,\eta\right),
\end{align}
$p_{\rm in}$ denoting the initial filler concentration of the preform. Employing only steady solutions, we can substitute the time integration by a spatial integration following the trajectory $\gamma(r_f)$ of a material element during drawing. The trajectory is hereby identified by the final radial position of the material element $r_f$ and it can be parametrized using the axial position $z$, with the radial position being a function of $z$. Equation~(\ref{eq:pe_sol}) can then be rewritten as
\begin{align}
\label{eq:pe_sol_z}
p_e = p_c + (p_{\rm in} - p_c)\,\exp\left(-a\int_{\gamma(r_f)}\frac{dz}{v}\,\sqrt{II_D}^{\,m}\eta\right).
\end{align}\par

The conductivity is then determined using a common percolation law,
\begin{align}
\label{eq:perc}
\sigma = \sigma_{\rm in}\,\left(\frac{p_e-p_c}{p_{\rm in}-p_c}\right)^c,
\end{align}
with initial conductivity of the preform $\sigma_{\rm in}$ and critical exponent $c$. Substituting Eq.~(\ref{eq:pe_sol_z}) in (\ref{eq:perc}) then leads to
\begin{subequations}
\label{eq:cond_distr}
\begin{align}
\label{eq:cond_distr_short}
\sigma(r_f) &= \sigma_{\rm in}\exp\left(-\,a\,c\,\int_{\gamma(r_f)}\frac{dz}{v}\,\sqrt{II_D}^{\,m}\eta\right)
\end{align}
\end{subequations}
for the final conductivity of the fully drawn fiber. The initial conductivity of the undeformed material $\sigma_{\rm in}$ and the exponent $a\,c$ are determined by fitting the experimental data, similar to the work of \citet{saphiannikova_superposition_2012}.

\section{Materials and Methods}
\subsection*{Materials}
Polysulfone and polycarbonate plates are purchased from Boedecker Plastics, Inc. Pellets of polycarbonate and a polycarbonate/carbon nanotube masterbatch ($15\,\rm{wt\%}$) are purchased from Nanocyl SA and melt-mixed together in a DSM twin-screw micro-compounder at $240\,\rm{^\circ C}$ to achieve a concentration of $3\,\rm{wt\%}$; the screw rotation speed is set to $100\,{\rm rpm}$ and the mixing time is $5\,\rm{min}$; the extruded material is then hot pressed into a sheet at $200\,^\circ\rm{C}$ for 2 minutes with a Lauffer press. The polycarbonate/carbon black nanocomposite is purchased from Westlake Plastics in sheet form ($0.25\,\rm{mm}$ thick). Metallic ribbons ($2\,\rm{mm}$ x $1\,\rm{mm}$) of low melting point tin-zinc and bismuth-tin alloys are purchased from Indium Corporation.
\subsection*{Preform fabrication}
The polysulfone or polycarbonate plates are cut to the dimensions of the metallic consolidation mold ($170\,\rm{mm}$ x $24\,\rm{mm}$) employed and milled with grooves dimensioned to fit the nanocomposite sheets and metallic electrodes. After positioning the materials together, they are thermally welded by hot pressing the preform in the consolidation mold in a Lauffer press. A pressure of $1\,\rm{N/cm^2}$ is applied for $1\,\rm h$ at a temperature of $220\,\rm{^\circ C}$ ($185\,\rm{^\circ C}$) for preforms with a polysulfone (polycarbonate) cladding.
\subsection*{Thermal drawing}
The preforms are thermally drawn into fibers in a custom-made draw tower. The temperatures of the three heating zones of the tube furnace are typically set to $150/260/100\, \rm{^\circ C}$ for polycarbonate and $200/320/120\,\rm{^\circ C}$ for polysulfone. The feeding speed is 2 mm/s, and the drawing speed is varied depending on the targeted draw ratio.
\subsection*{Conductivity measurements}
The fibers have metallic electrodes, which are in contact with the nanocomposites. The cladding above the metallic electrodes is locally polished away to enable a measurement of the electrical resistance perpendicular to the drawing direction of the nanocomposite layer between two electrodes, using a Keithley 2450 Sourcemeter. From the resistance measurement, the conductivity is then calculated using the geometrical dimensions of the region considered.

\section{Parameter determination}
\begin{table}
\begin{tabular}{lllll}
\hline
Material & $\rho / (\rm kg\,m^{-3})$ & $c_p / (\rm J\,kg^{-1}\,K^{-1}$ & $k / (\rm W\,m^{-1}\,K^{-1})$ & $E_a / (\rm kJ\,mol^{-1})$\\
 \hline
 PSu & $1240$ & $1000$ & $0.25$ & $219.8$\\
 PC & $1200$ & $1800$ & $0.25$ & $182.0$\\
 \hline
\end{tabular}
\caption{Material parameter values of PSu and PC used in this study. Apart from the activation energy, literature values are used.}
\label{tab:matparam}
\end{table}
\begin{table}[t]
\begin{tabular}{lllll}
\hline
Material & $h / (\rm W\,m^{-2}\,K^{-1})$ & $\Lambda$ & $\Delta$ & $T_{\rm max} / \rm ^\circ C$\\
 \hline
 PSu & $20$ & $0.22$ & $0.5$ & $225.7$\\
 PC & $30$ & $0.17$ & $0.45$ & $182.9$\\
 \hline
\end{tabular}
\caption{Parameter values determined by fitting the fiber radius profile.}
\label{tab:fitparam}
\end{table}

\begin{figure}
\centering
\includegraphics[width=0.65\columnwidth]{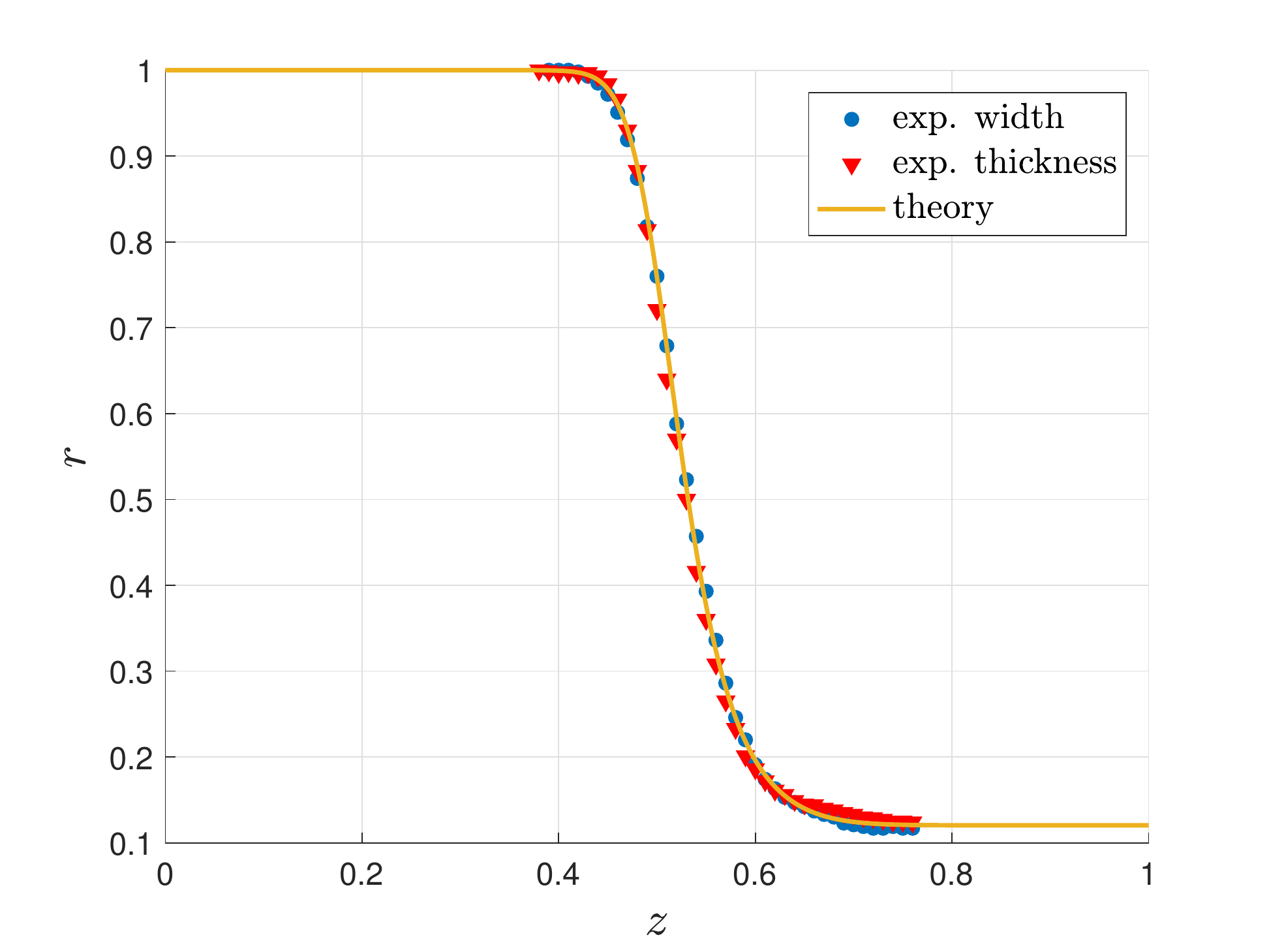}
\caption{Fit of predicted dimensionless fiber radius profile to the experimentally determined fiber width and thickness profiles for a Polysulfone fiber. The draw ratio is $Dr=8.31$ and the parameters obtained from the fit are $\Lambda = 0.5$, $\Delta = 0.22$, $h=20\,\rm{W\,m^{-2}\,K^{-1}}$ and $T_{\rm max} = 225.7\,^\circ \rm C$.}
\label{fig:shape_fit}
\end{figure}

Table~\ref{tab:matparam} lists the values of the material parameters used in this study for both PSu and PC. The activation energy is determined by shear rheometry, while the remaining parameters are obtained from the material specification sheet of the supplier and Comsol Multiphysics material library. The effective heat transfer coefficient $h$ and the effective air temperature profile parameters $\Lambda$, $\Delta$ and $T_{\rm max}$, the latter being necessary for calculating $\mu$, are determined by fitting the predicted fiber radius profile to the experimental observation. In addition to that, we make sure that the experimentally measurable tension necessary to draw the fiber is reproduced by the theoretical prediction. As the fabricated fibers have a rectangular cross-section, we use both the dimensionless width and thickness profiles for fitting. Both profiles exhibit a very similar dimensionless form, which comforts the use of an axisymmetric model. An example of such a fit is given by Fig.~\ref{fig:shape_fit} and the parameters determined for both cladding materials are given in Table~\ref{tab:fitparam}.\par

\section{Evolution of temperature, viscosity and velocity}
\begin{figure}
\centering
\includegraphics[width=0.65\columnwidth]{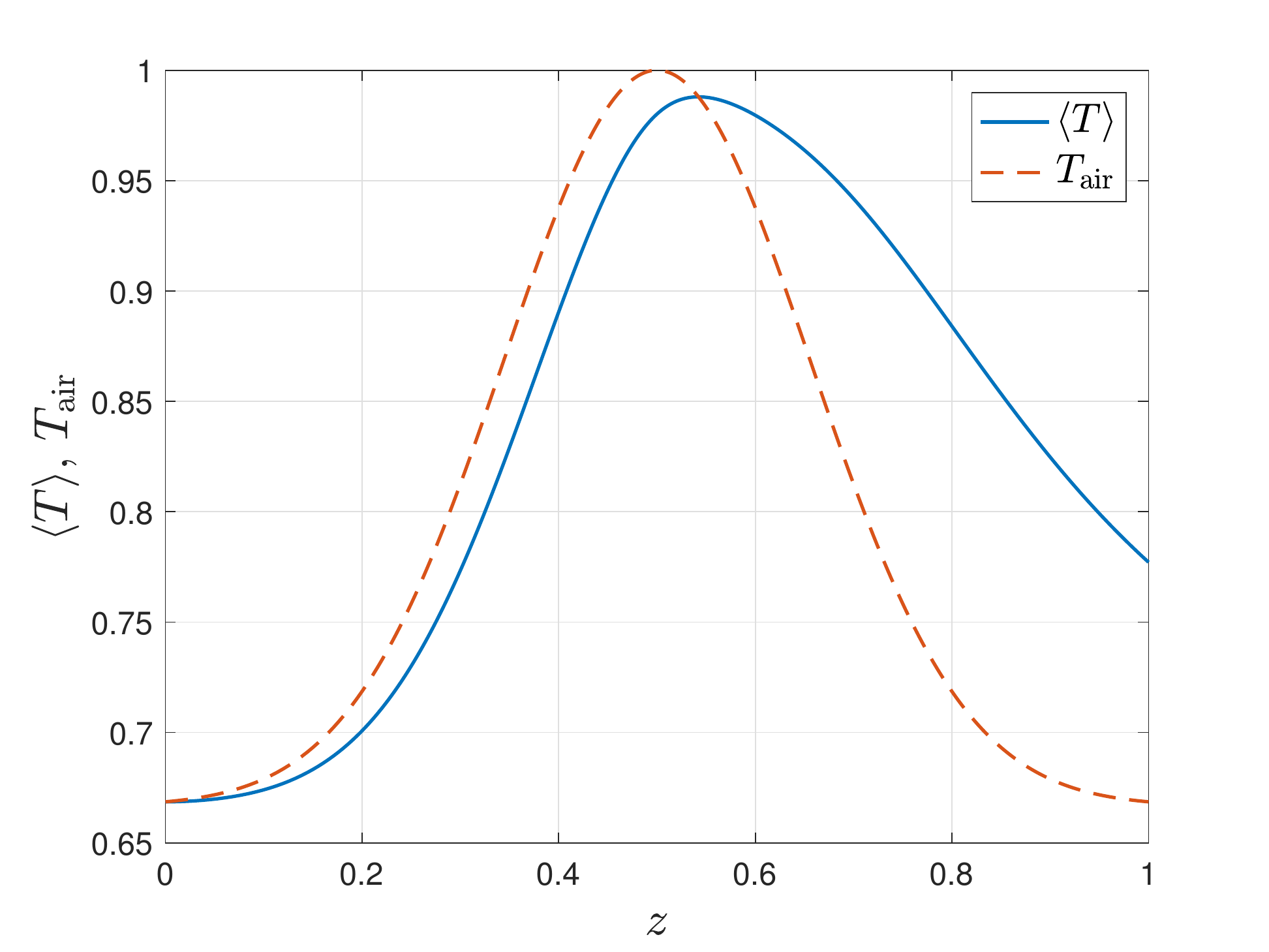}
\caption{Comparison between cross-section averaged fiber temperature profile $\langle T\rangle$ and air temperature $T_{\rm air}$ for a PSu fiber drawn at $Dr=8.31$.}
\label{fig:T_prof}
\end{figure}
\begin{figure}[h!p]
\centering
\includegraphics[width=0.9\columnwidth]{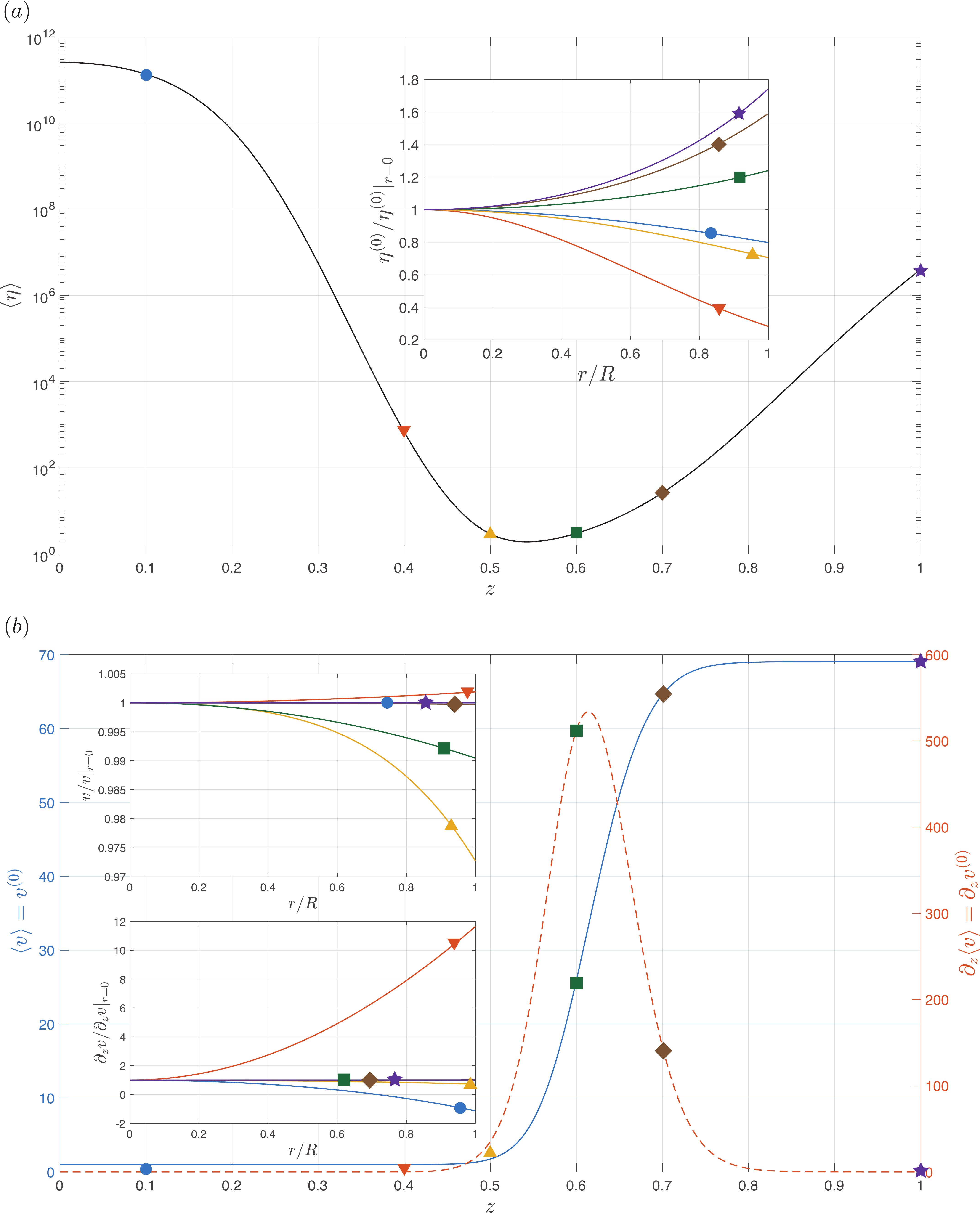}
\caption{Axial profile of (a) cross-section averaged viscosity $\langle \eta\rangle$ and (b) axial velocity $\langle v\rangle$ and strain rate $\partial_z\langle v\rangle$ for a PSu fiber drawn at $Dr=8.31$. The insets show the radial distribution of the non-averaged quantities at several positions on the $z$-axis.}
\label{fig:eta_v_details}
\end{figure}
We provide here additional information on the axial and radial shape of temperature, viscosity and axial velocity. Figure~\ref{fig:T_prof} compares the air temperature profile to the cross-section averaged temperature in a PSu fiber. While the maximum temperature $T_{\rm max}$ is almost reached by the material, justifying the choice of it as reference scale, the material temperature exhibits a delayed reaction to the air temperature, which is especially visible in the cooling part ($z>0.5$). The delay is primarily determined by the Stanton number, which is $St=32$ here, as it compares the heat transfer between the fiber and the surrounding to the internal convection. A higher $St$ thus means $\langle T\rangle$ approaching $T_{\rm air}$ closer.\par
Figure~\ref{fig:eta_v_details} shows the evolution of cross-section averaged viscosity, axial velocity and strain rate along the drawing axis. Similar to the temperature, the viscosity does not reach its initial value at $z=1$, but is six orders of magnitude higher than its lowest value around the middle of the drawing distance. As can be seen in the velocity and strain rate profiles, the fiber deformation is already finished at an earlier position and most of the deformation occurs between $z=0.4$ and $0.8$.\par
The insets in Fig.~\ref{fig:eta_v_details} show the radial distribution of the non-averaged quantities at several positions on the drawing axis. While viscosity and strain rate exhibit significant changes along the radius over the entire drawing length, the velocity varies only in a narrow range from the averaged value.